\documentclass[A4Paper,usenatbib,useAMS]{mnras}
\usepackage{multirow}
\usepackage{rotating}
\usepackage{pdflscape}
\usepackage{hyperref}
\usepackage{supertabular}
\usepackage[table]{xcolor}

\title[Diverse stellar haloes]{Diverse Stellar Haloes in Nearby Milky Way-Mass Disc Galaxies}

\author[Harmsen et al.]{Benjamin Harmsen$^{1}$, \thanks{Contact e-mail:\href{mailto:benharms@umich.edu}{benharms@umich.edu}}
						Antonela Monachesi$^{2}$, \thanks{Contact e-mail:\href{mailto:antonela@mpa-garching.mpg.de}{antonela@mpa-garching.mpg.de}}
						Eric F.\ Bell$^{1}$, \thanks{Contact e-mail:\href{mailto:ericbell@umich.edu}{ericbell@umich.edu}}
						Roelof S.\ de Jong$^{3}$, 
                        \newauthor
						Jeremy Bailin$^{4,5}$, 
						David J.\ Radburn-Smith$^{6}$, 
						Benne W.\ Holwerda$^{7}$ 
\\
$^{1}$University of Michigan, Department of Astronomy, 311 West Hall, 1085 South University Ave., Ann Arbor, MI 48109-1107\\
$^{2}$Max Planck Institut f{\"u}r Astrophysik, Karl-Schwarzschild-Str. 1, Postfach 1317, D-85741 Garching, Germany\\
$^{3}$Leibniz-Institut f\"ur Astrophysik Potsdam (AIP) An der Sternwarte 16, 14482 Potsdam, Germany\\
$^{4}$Department of Physics and Astronomy, University of Alabama, Box 870324, Tuscaloosa, AL 35487-0324, USA\\
$^{5}$National Radio Astronomy Observatory, P.O. Box 2, Green Bank, WV, 24944, USA\\
$^{6}$Department of Astronomy, University of Washington, 3910 15th Ave NE, Seattle, WA 98195,USA\\
$^{7}$Leiden Observatory, Niels Bohrweg 2, NL-2300 CA, Leiden, The Netherlands\\
}

\begin{document}
\label{firstpage}
\pagerange{\pageref{firstpage}--\pageref{lastpage}}
\maketitle





\begin{abstract}
We have examined the resolved stellar populations at large galactocentric distances along the minor axis (from 10\,kpc up to between 40 and 75\,kpc), with limited major axis coverage, of six nearby highly-inclined Milky Way-mass disc galaxies using HST data from the GHOSTS survey. We select red giant branch stars to derive stellar halo density profiles. The projected minor axis density profiles can be approximated by power laws with projected slopes of between $-2$ and $-3.7$ and a diversity of stellar halo masses of $1-6\times 10^{9}M_{\odot}$, or $2-14\%$ of the total galaxy stellar masses. The typical intrinsic scatter around a smooth power law fit is $0.05-0.1$ dex owing to substructure. By comparing the minor and major axis profiles, we infer projected axis ratios $c/a$ at $\sim 25$\,kpc between $0.4-0.75$. The GHOSTS stellar haloes are diverse, lying between the extremes charted out by the (rather atypical) haloes of the Milky Way and M31. We find a strong correlation between the stellar halo metallicities and the stellar halo  masses. We compare our results{ with cosmological models}, finding good agreement between our observations and accretion-only models where the stellar haloes are formed by the disruption of dwarf satellites. In particular, the strong observed correlation between stellar halo metallicity and mass is naturally reproduced. Low-resolution hydrodynamical models have unrealistically high stellar halo masses.{ Current high-resolution hydrodynamical models appear to predict stellar halo masses somewhat higher than observed but with reasonable metallicities, metallicity gradients and density profiles.} 

\end{abstract}


\section{Introduction}

The current favoured cosmological model $\Lambda$-Cold Dark Matter ($\Lambda$CDM) is hierarchical, predicting that dark matter haloes are assembled over
time through the collisionless accretion and mergers of smaller haloes. Stars form in the centres of the larger dark matter haloes \citep{White_Rees78,Moore1999}, where the number of stars that form in low mass haloes is dramatically suppressed compared to larger haloes (likely owing to feedback from supernovae; e.g., \citealp{DekelSilk1986}; \citealp{Cole1991}; \citealp{Wheeler14}). As satellite haloes merge with the main halo, their (typically meager) stellar components tidally disrupt and are spread into a diffuse and structured stellar halo \citep{Bullock01,BJ05,Cooper10}. The resulting stellar haloes are expected to exhibit steep density profiles, have abundant substructure and show considerable halo-to-halo variation in their properties, all of which are expected to be closely tied with their merger histories. The goal of this work is to carefully characterise the density profiles, projected axis ratios, stellar masses and substructure of the stellar haloes of six nearby roughly Milky Way mass disc galaxies with resolved stellar population measurements from the Hubble Space Telescope \citep{RadSmith11,M16a}.

While models in which stellar haloes are composed of the tidal debris from dwarf galaxy disruption alone share a number of broad qualitative predictions --- diffuse, centrally concentrated, highly structured stellar haloes whose metallicities reflect the metallicities of the disrupted dwarf galaxies --- some of the quantitative predictions for stellar halo properties vary considerably from model to model. Stellar halo masses and metallicities of Milky Way-mass galaxies vary considerably more from halo to halo in the \citet{Cooper10} and \citet{Gomez12} models than in \citet{BJ05}, likely from a wider range of satellite accretion histories (the importance of input satellite metallicity distributions is emphasised by \citealp{Tumlinson10}). Stellar haloes in the $N$-body only models of \citet{Cooper10} are triaxial, whereas haloes in \citet{BJ05} are oblate. \citet{Bailin14} show that oblate haloes are the natural result of the growth of a stellar halo in a potential where the baryons are allowed to grow into a galaxy with a prominent disc, {suggesting that the presence of a disc potential (incorporated in \citealp{BJ05} but absent in \citealp{Cooper10}) is a key driver of stellar halo oblateness}. Furthermore, while the \citet{BJ05} models all lack a significant metallicity gradient, the metallicity gradients of the \citet{Cooper10} models vary considerably, from no gradient to relatively rapid changes in metallicity with radius. This diversity in model predictions signals the strength of observations to test and guide the models.

In addition to the stars accreted from disrupted satellites, the inner
parts of stellar haloes are predicted to have a considerable population of in-situ stars formed in the main galaxy potential \citep{Zolotov09, Font11, Tissera14, Pillepich15}. The physical ingredients of such models have significant uncertainties, e.g., the modeling of feedback processes, stellar winds, star formation recipes, and gas dynamics; all of which have significant impact on how stars populate dwarf satellites, the shape of the potential of the main galaxy, the fraction of in-situ stars, and even whether the in-situ stars are a common feature of all haloes \citep{Bailin14}. The mass and extent of expected in-situ haloes are predicted to vary by large factors, ranging from being dominant at radii of even 30 kpc \citep[e.g.,][]{Font11} to being dominant only at $< 5$ kpc \citep[e.g.,][]{Pillepich15}. The prominence of in-situ stars is expected to be a function of position with respect to the disc (more prominent along the major axis, and less detectable along the minor axis; \citealt{M16b}), galaxy mass, and merger history \citep{Zolotov11}.

All of these considerations motivate the careful characterisation of a sizeable sample of stellar haloes. Yet, owing to the observational challenge of detecting low surface brightness and diffuse features in more distant galaxies, the stellar populations and shapes of the main body of the haloes of only the Milky Way and Andromeda have been studied in-depth to date (e.g., \citealp{Ivesic00}, \citealp{Newberg07}, \citealp{Bell08}, \citealp{Gilbert12}, \citealp{Deason13}, \citealp{Gilbert14}, \citealp{Ibata14}). While both haloes are richly substructured and qualitatively agree with the $\Lambda$CDM paradigm of galaxy formation, they display significant differences. The Milky Way halo has a weak to no metallicity gradient \citep{Sesar11,Xue15} and its stellar density distribution can be described by a broken power-law --- within $25-30$ kpc, it follows an oblate, $\rho \propto r^{-\gamma}$ power-law distribution with index $\gamma \sim2.5-3$ \citep{Yanny00,Bell08,Juric08,vanVledder16} whereas a more rapidly declining stellar density is detected beyond $\sim 30$ kpc, with $\gamma \sim 3.5$ \citep{Deason11,Sesar11,Deason14,Cohen15,Slater16}. M31, on the other hand, has a clear metallicity gradient with a 1 dex variation in $[\rm{Fe/H}]$ from 10 to $\sim$100 kpc \citep{Gilbert14,Ibata14} and its stellar density distribution can be described by a single power-law with $\gamma \sim3.3$  \citep{Guhathakurta05,Gilbert12}. In order to test model predictions and quantify the halo-to-halo variations such as differences in metallicity profiles, fraction of stellar halo created in-situ and accreted, stellar halo morphology, etc.,  we need to observe the stellar halo properties of more similar mass galaxies. In particular, the stellar halo density profiles and shapes can provide important constraints on the merging and accretion history of a galaxy \citep{Johnston08,Deason13, Amorisco15}.

Over the last decades, a number of efforts have sought to characterise the diffuse stellar envelopes around galaxies. 
Integrated light studies of nearby galaxies show that stellar streams (thought to be from the disruption of dwarf galaxies) are reasonably common  \citep{MalinHadley97,Shang98,Mihos05,Tal09,MDelgado10,Paudel13,Watkins15,Merritt16}. 
Such studies are challenging, requiring excellent control of scattered light \citep{Slater09,AbrahamVD14}. While it is often possible to control scattered light well enough to uncover tidal streams as local enhancements in surface brightness, it is extremely challenging to control scattered light well enough to convincingly and correctly recover the brightness profile of the larger scale `aggregate' (sometimes, somewhat misleadingly termed `smooth') stellar halo {(\citealp{deJong08a}, although see \citealp{Dsouza14} and \citealp{Merritt16} for encouraging progress)}. Such broad scale and diffuse structures are possible (with substantial observational cost) to detect and characterise by resolving individual (typically red giant) stars in nearby galaxy stellar haloes. Using such methods, diffuse stellar haloes have been detected and characterised around a number of nearby galaxies (\citealp{Mouhcine05b}, \citealp{M16a}, M81: \citealp{Barker09}, \citealp{Mon13}, NGC 253: \citealp{Bailin11}, \citealp{Greggio14}; NGC 891: \citealp{Mouhcine10}; Cen A: \citealp{HH02}, \citealp{Rejkuba14}, \citealp{Crnojevic16}; NGC 3115: \citealp{Peacock15}; NGC 3379: \citealp{Harris3379}; NGC 3377: \citealp{Harris3377}). Yet, quantitative analysis of the density profiles has proven challenging. For the disc-dominated galaxies that we focus on in this paper, such analyses were carried out only for NGC 253 \citep{Bailin11,Greggio14} and M81 \citep{Barker09} where flattened ($0.4<c/a<0.6$), steeply declining power law density profiles were determined. In addition, in common with the integrated light studies, substantial substructure in stellar haloes (streams or shells) has been uncovered in many cases (e.g. \citealp{Bailin11}, \citealp{Greggio14}, \citealp{Crnojevic16}). These efforts illuminate the path towards quantifying halo properties with resolved stellar populations, but have not yet yielded a sizeable sample of galaxies with quantified stellar halo properties. 

In this paper we present the projected red giant branch star density profiles out to projected radii $\sim 40-75$ kpc along the disc's minor axis and major axis profiles out to smaller radii of the stellar haloes of six massive nearby disc galaxies using Hubble Space Telescope (HST) observations from the Galaxy haloes, Outer discs, Substructure, Thick discs, and Star clusters (GHOSTS) survey \citep{RadSmith11, M16a}. We focus on the massive galaxy subset of the GHOSTS survey both because their stellar haloes are prominent and straightforward to characterise in our dataset and because many models of stellar halo properties focus on galaxies in this stellar mass range ($4 \times 10^{10} M_{\odot} < M_* < 8 \times 10^{10} M_{\odot}$). With these data, we estimate the stellar halo density profiles, degree of substructure, projected axis ratio, and halo stellar masses of this sample, and can combine these measurements with stellar halo population measurements from \cite{Mon13} and \citet{M16a} to explore correlations between stellar halo structures and stellar populations. Section~\ref{sec2} provides an overview of the GHOSTS survey. The data reduction and photometry are summarised in Section~\ref{sec3}.  We present our results for the red giant branch star density profiles, surface brightnesses, estimated shapes, degree of substructure and stellar halo masses in Section~\ref{sec4}. We build intuition about how results from our sparsely-sampled data might compare with global halo properties using models in Section~\ref{BJ_Test}. With this intuition in hand, we discuss the results for each galaxy in more depth and compare with previous results in Section~\ref{indiv}. In Section~\ref{disc}, we explore the correlations between stellar halo and galaxy properties and compare with theoretical predictions{; readers interested primarily in the big picture results are invited to skip directly to Section~\ref{disc}}. We summarise and conclude in Section~\ref{sec6}. 

\section{Observations: The GHOSTS survey}
\label{sec2}

The Galaxy haloes, Outer discs, Substructure, Thick discs, and Star
clusters (GHOSTS) survey, \citep{RadSmith11} is a large HST program designed to resolve the stars in the outskirts of eighteen Local Volume disc galaxies of different
masses, luminosities, and inclinations --- the largest such study to date. Fields along the principal axes of each galaxy were observed, reaching projected galactocentric distances as large as $\sim75$ kpc. GHOSTS observations provide star counts and colour-magnitude diagrams (CMDs) reaching typically $\sim 2-3$ magnitudes below the tip of the red giant branch (TRGB). Using the RGB stars as tracers of the stellar halo population, we are able to study the size and shape of each stellar halo as well as the properties of their stellar populations such as
age and metallicity.
A more detailed description of the survey can be found in \cite{RadSmith11} and \cite{M16a}.
 
We selected six galaxies for study in this paper. These galaxies were chosen because they are the most massive among those in GHOSTS,  comparable in stellar mass and rotation velocity to the Milky Way. The galaxies are also highly inclined or edge-on, which ensures minimal to no disc contamination when observing along the minor axis beyond $10$ kpc.

\begin{figure*}\centering
\includegraphics[width=170mm]{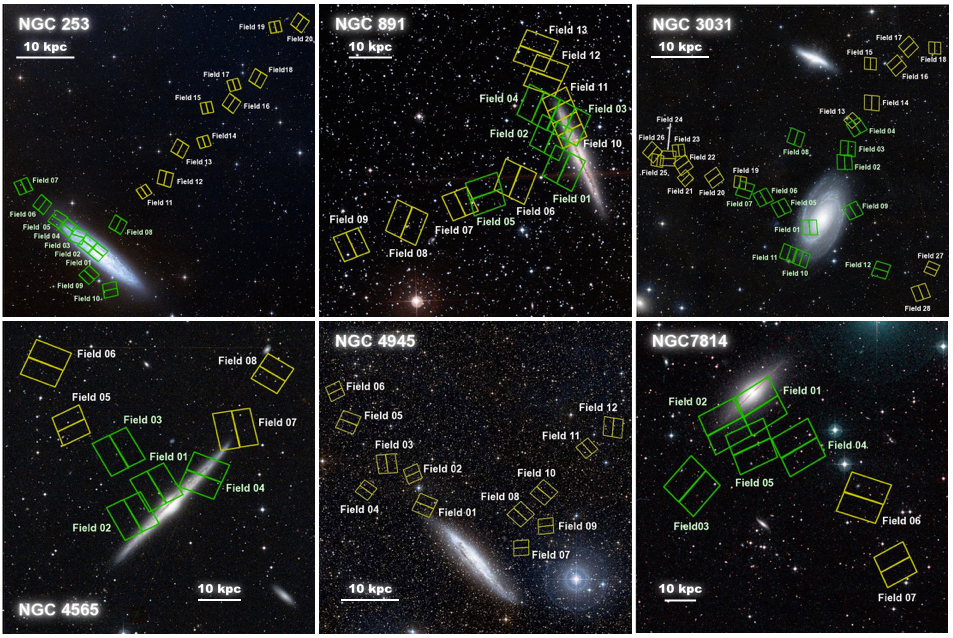}
\caption{Location of the GHOSTS HST ACS/WFC and WFC3/UVIS fields overlaid on DSS coloured images of each galaxy. Green fields were introduced in \citet{RadSmith11} whereas yellow fields were presented by \citet{M16a}. North is up and east is to the left. Fields were placed mostly along the principal axes with some at intermediate position angles. This strategy allows us to both probe their haloes out to projected distances of  $R \sim 40 -75$\,kpc along the minor axis from the galactic centre as well as to measure the halo structure and stellar population differences where different regions are observed. For our purposes, not all GHOSTS fields were used as some lie along intermediate axes or are too close to the disc; the list of fields that we have analyzed here is given in Table 1 of \protect\citet{M16a}.}
\label{fields}
\end{figure*}

Fields were chosen to lie on the galactic discs \citep[to study disc structure, dust and flaring; e.g.,][]{deJong07,RS12,RS14,Streich16} or much further out along the major and minor axes of the main body of the galaxy \citep[to study outer discs and stellar haloes;][]{Mon13,M16a}, with a few pointings exploring intermediate position angles. In practice, the outermost major axis fields have stellar populations and structures indicative of being dominated by stellar halo, permitting the projected shape of the stellar halo to be estimated if one assumes alignment between the principal axes of the disc and halo \citep[we examine this assumption later in \S\protect\ref{BJ_Test}]{BJ05,Bailin14,Pillepich15}. The locations of the fields extend out to distances of $\sim$40-75\,kpc from the centre of the galaxy along the minor axis, depending on the galaxy, as shown in Figure~\ref{fields}.

\section{Data reduction and Photometry}
\label{sec3}

We summarise in this section the main data reduction steps and
 stellar photometry performed for each exposure using the
 GHOSTS pipeline. We refer the interested reader to \citet{RadSmith11}
 and \citet{M16a} where the
 pipeline for the data are described for HST/ACS and HST/WFC3 respectively.
 
 We downloaded the ACS $*$\verb+_flc+ FITS images from the Hubble Data
Archive MAST\footnote{\url{http://archive.stsci.edu}}, which have been bias-subtracted, cosmic
ray flagged and removed, 
flat fielded and corrected for charge transfer efficiency \citep[CTE;][]{Anderson_bedin10}.  For the WFC3 images, we have generated the
$*$\verb+_flc+ FITS images locally from the $*$\verb+_raw+ FITS images
downloaded from MAST, using a code provided by STScI, since the pixel
based CTE correction is not yet a part of the WFC3/UVIS pipeline.  For
each field, we combine the individual FLC images per filter using
AstroDrizzle \citep{Gonzaga12}. The resulting image per field and
filter is a drizzled DRC FITS image, which has been corrected for
geometric distortion.

We used DOLPHOT, an updated version of HSTphot \citep{Dolphin00}
for ACS and WFC3 images, to perform simultaneous point-spread function (PSF)
fitting photometry on all the individual FLC exposures per field. The DOLPHOT parameters used for the GHOSTS fields are
given in Table A2 of \citet{M16a}. 
DOLPHOT provides the position of each star relative to the $F814W$
drizzled image, together with the instrumental HST magnitudes in the VEGAmag system already
corrected for CTE loss and with aperture corrections calculated using
isolated stars. The DOLPHOT output includes various diagnostic
parameters that are used to discriminate between 
PSF-like stars and non-PSF shaped detections such as cosmic
rays and background galaxies.

When attempting to measure the number of faint stars in sparsely-populated (with tens to hundreds of stars) HST fields, compact background galaxies are the most important source of contamination.  We impose several selection criteria to the ACS and WFC3 catalogs, termed ``culls'' by \citet{RadSmith11} and \citet{M16a},
using diagnostic parameters
such as sharpness and crowding to distinguish between PSF-shaped sources and sources more or less extended than the PSF. These culls were 
applied to the photometry
output, which removed $\sim$95\% of the contaminants. The different
culls and details on how they were obtained for the ACS and WFC3 data
can be found in \citet{RadSmith11} and \citet{M16a}
respectively. In addition, we
used SE\verb+XTRACTOR+  \citep{Bertin_arnouts96} to construct a mask for all
extended sources for each field, which include both background
galaxies as well as bright foreground Milky Way (MW) stars. Detected
sources that fall in the masks were removed from the photometry output
file.
The shorter observations of the WFC3 fields of our closest galaxies (all WFC3 fields in NGC 3031 and NGC 4945 as well as Field 14 in NGC 253) have only one exposure in the $F606W-$band image. Because our pipeline was unable to remove the cosmic rays in these single exposure $F606W$ images, many cosmic rays, which are as compact or more compact than real stars, remain in the $F606W$ images.  Following \citet{M16a}, we performed an iterative analysis where objects were detected in the $F606W$ and $F814W$ images; those objects which are much too bright in $F606W$ to be real stars were masked out and the photometry recomputed. These masked cosmic rays were added to the SEXTRACTOR mask, and were used to reject spurious sources. 

We note that contamination from MW foreground stars was not
significant within the colour and magnitude range of interest for four of our six sample galaxies at high galactic latitude. 
Foreground contamination was more severe for
NGC 4945 and NGC 0891 fields since these galaxies are at a low
galactic latitude. 
Based on the CMDs and colour distributions of fields simulated by TRILEGAL\footnote{\url{http://stev.oapd.inaf.it/cgi-bin/trilegal} } \citep{Girardi05} and Besan\c con\footnote{\url{http://model.obs-besancon.fr/}} \citep{Robin03} models, we adopted a colour cut for NGC 4945 CMDs to remove most MW contaminants. 

In order to assess the completeness of our data and to quantify the
photometric uncertainties, we have performed extensive
artificial star tests (ASTs) on each exposure, as described 
by \citet{RadSmith11}. Approximately 2,000,000 fake stars
were injected in each exposure with realistic colours and magnitudes
and distributed such that they follow the observed stellar density
gradient. We run DOLPHOT on each fake star at a time and we applied
the same culls as in the real output photometry
catalogs. Artificial stars that did not pass the culls were considered to be
lost. The completeness level was calculated as the ratio of
recovered-to-injected number of artificial stars at a given colour and
magnitude bin.

\begin{figure*}\centering	
	\includegraphics[width=170mm]{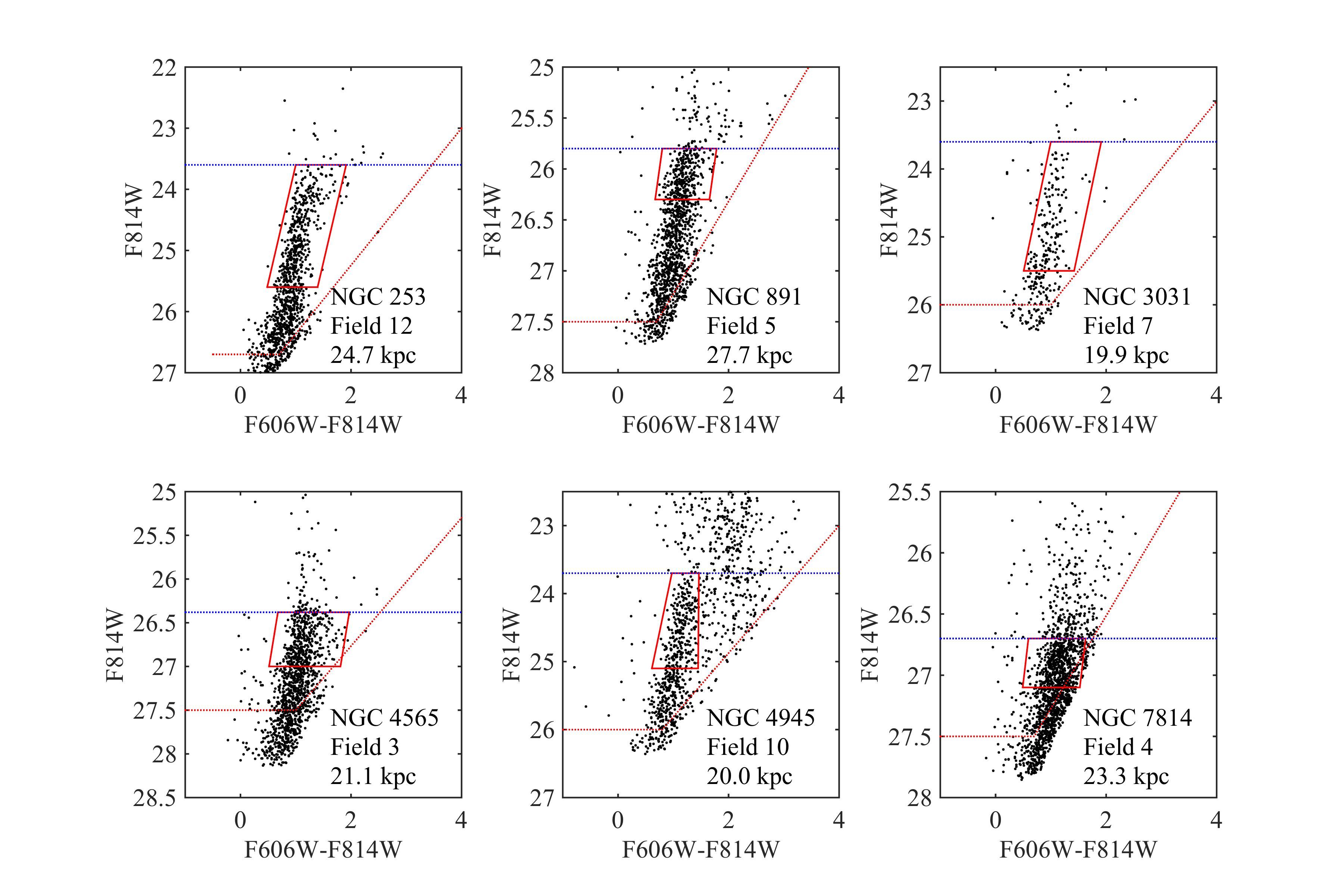}
	\caption{Representative F606W-F814W vs.\ F814W GHOSTS colour-magnitude diagrams for each of our six target galaxies. The red boxes represent the selection cuts (see Section~\ref{sec4}).  Only detections selected to be stars following the \protect\citet{RadSmith11} and \protect\citet{M16a} photometric culls are shown. The TRGB is indicated by a dotted blue line; the 50\% completeness limits, as determined from ASTs, are represented by a dotted red line. The field number as well as their projected radial distance from the galaxy centre are indicated in each panel.}
	\label{CMDs}
\end{figure*}

Examples of a resulting CMD per galaxy,
after the masks and culls were applied, are shown in
Figure~\ref{CMDs}.  The mean distance of each field from the galaxy centre
is indicated in each panel. These CMDs are largely free from
background and foreground sources. The 50\% completeness level of each
field is indicated with a dashed red line.  Several
CMDs for each galaxy are shown in \citet{RadSmith11} and \citet{M16a}. All the CMDs for the entire survey can be found in the
GHOSTS website at \url{http://vo.aip.de/ghosts/}.

\section{Results}
\label{sec4}

In this section, we present the methods used to calculate and fit the red giant branch density profiles of the six galaxies in our sample, and the results of those fits. More detailed discussion of the fits on a galaxy-by-galaxy basis, along with tests of our methods and comparison with models is presented later in Sections \ref{indiv} and \ref{disc}.

\subsection{Stellar Density Profiles}

In order to characterise the stellar haloes of our six sample galaxies, we choose to select and analyze stars with the colours and magnitudes of relatively metal-poor red giant branch (RGB) stars at the distances of each of the target galaxies. As tracers of the stellar halo, RGB stars offer a number of advantages. The RGB is a prominent feature of the CMD of essentially all intermediate-age and old stellar populations. RGB stars are relatively numerous, offering a large sample of stars to characterise and study. They also have a well-defined maximum luminosity \citep[e.g.,][]{Bellazzini01}, allowing measurement of the stellar halo distances and an important check that the stars under consideration indeed belong to the target galaxy. RGB stars of the metallicities and ages thought to dominate the bulk of stellar halo populations, with ages in excess of a few billion years and metallicities below 1/3 solar, have a moderately well-defined range of colours, making their identification and characterisation relatively straightforward. Finally, RGB star colours do vary somewhat as a function of population parameters (primarily metallicity, see e.g., \citealt{Hoyle_Schwarzschild55,Sandage_smith66}), offering insight into the stellar populations of the target stellar haloes. 

Candidate RGB stars were selected by making cuts in colour-magnitude space.  We select candidate RGB stars to have magnitudes between the TRGB, as presented in \citet[see their Table C1]{M16a}, and a limit chosen to lie above the $50-70$\% completeness limits as determined by the results of the ASTs; this limit is between 0.5 and 1.5 magnitudes fainter than the TRGB. In practice, this limit depends primarily on the distance to the galaxy (which set our depth compared to the TRGB), where more distant galaxies tend to have shallower CMDs and therefore smaller magnitude ranges for RGB star selection. The colour limits at the blue end are designed to prevent contamination from Main Sequence and Helium-Burning stars\footnote{In NGC 3031's case, this meant a blue colour limit that is very close to the RGB (see Fig.\ \protect\ref{CMDs}, as there are substantial numbers of blue stars in M81's outskirts; \citealp{Okamoto15}).}, while the colour limits at the red end are designed to prevent contamination from very metal-rich disc or Milky Way foreground stars as well as to ensure the $50-70$\% completeness level of the selected stars. In all cases, the colour selection encompasses the vast majority of halo stars at all relevant radii (minor axis radii $>5$kpc and major axis radii $>20$kpc). A representative CMD for each galaxy is presented in Fig.\ \ref{CMDs} along with the adopted selection cuts. For a given galaxy the same RGB selection cuts were used for all fields. Only stars inside the selection cuts were used to compute the stellar density profiles. 

Candidate RGB stars in each field were divided into bins based on their radial distance from the centre of the galaxy. The bins were chosen for each field to balance counting statistics on one hand with a fine enough radial sampling to allow detection of density gradients within a field and substructure, if it exists.  Due to the sparse nature of the outer fields, fewer bins were typically used at greater radial distances.  In order to minimise contamination by disc stars, we use only stars with radial distances greater than 5 kpc for the minor axis fields and 20 kpc for the major axis fields; the full list of fields that we analyze is given in Table 1 of \citet{M16a}. In each bin, the results from the ASTs were used to correct the star counts for photometric incompleteness. {NGC 7814 presents a unique case of severe crowding in the innermost fields.  This crowding results in significant undercounting of stars. Deep IRAC 3.6$\micron$ imaging from S4G \citep{MM15} detects extended light out to radii of $\sim$9\ and $\sim$23\,kpc along the minor and major axis, respectively, where our data are crowded; accordingly, we use those surface brightnesses as additional datapoints. Further description can be found in section \ref{7814Discussion}.}

To estimate the area in which stars can be reliably detected in each bin, we need to account for the regions of the images that are discarded. The mask generated using SEXTRACTOR (see Section~\ref{sec3}) was used to remove any detections near the locations of unresolved background galaxies, bright foreground stars, bad pixels or globular clusters.  Detections that fell within a certain distance (25 pixels for ACS and 5 pixels for WFC3) of any masked area were removed to ensure that our star catalogs are minimally contaminated by spurious detections in the vicinity of contaminating objects. 
The area of each bin was calculated by counting the pixels inside the previously determined radial bins and then subtracting out unused pixels from the mask. 

A major advantage of using resolved stellar populations to study low surface brightness stellar haloes is that the fore/background object counts correspond to a faint limiting surface brightness. Accordingly, the effects of fore/background subtraction are worth accounting for, but are only of modest importance. For most of our galaxies, our sparsest regions (typically in the outermost pointings) appear to have CMDs consistent with foreground Milky Way stars plus the few unresolved background galaxies left by the culls \citep{RadSmith11,M16a}. We choose to designate these areas as representing a fore/background density of RGB-coloured unresolved sources that should be subtracted from the area density for every pointing. For NGC 253, the outermost fields are well-populated by RGB stars, and we estimate (using the high-latitude control fields of \citealp{RadSmith11} and \citealp{M16a}) that $1/3 \pm 1/6$ of the RGB-coloured unresolved sources in the outermost fields (e.g., Field 20) are contaminants, and we adopt that density as an estimate of the fore/background density. For NGC 891, we adopt the number density of detections in the outermost Field 9 as the fore/background estimate. For NGC 3031 and NGC 4565, the lowest stellar density measurement was used as an estimate of the fore/background. For NGC 4945, the outermost fields appear to have a significant population of RGB stars, and we estimate that $1/2 \pm 1/4$ of Field 12's detected density is fore/background. For NGC 7814, Field 6 has a CMD consistent with mostly MW foreground stars, and is adopted as an estimate of fore/background. The uncertainty in the background was determined to be the square root of the number of stars except in the cases of NGC 253 and 4945, where we adopted an uncertainty of 50\% of the adopted background. For every RGB density measurement, the uncertainty on the background value was added in quadrature to the uncertainty for each data point. These corrections produce very modest effects on our final inferences. We have tested this by carrying out a full analysis without fore/background subtraction; all final measurements change by less than their quoted {\it random} error bars (as most of the inferences are driven by the higher surface brightness inner parts of the haloes), except for the minor axis power law slopes, which change by $\Delta \alpha \sim 0.1-0.5$, which is of the order of the systematic uncertainties in their power law slope.  

\begin{figure}\centering
	\includegraphics[width=80mm]{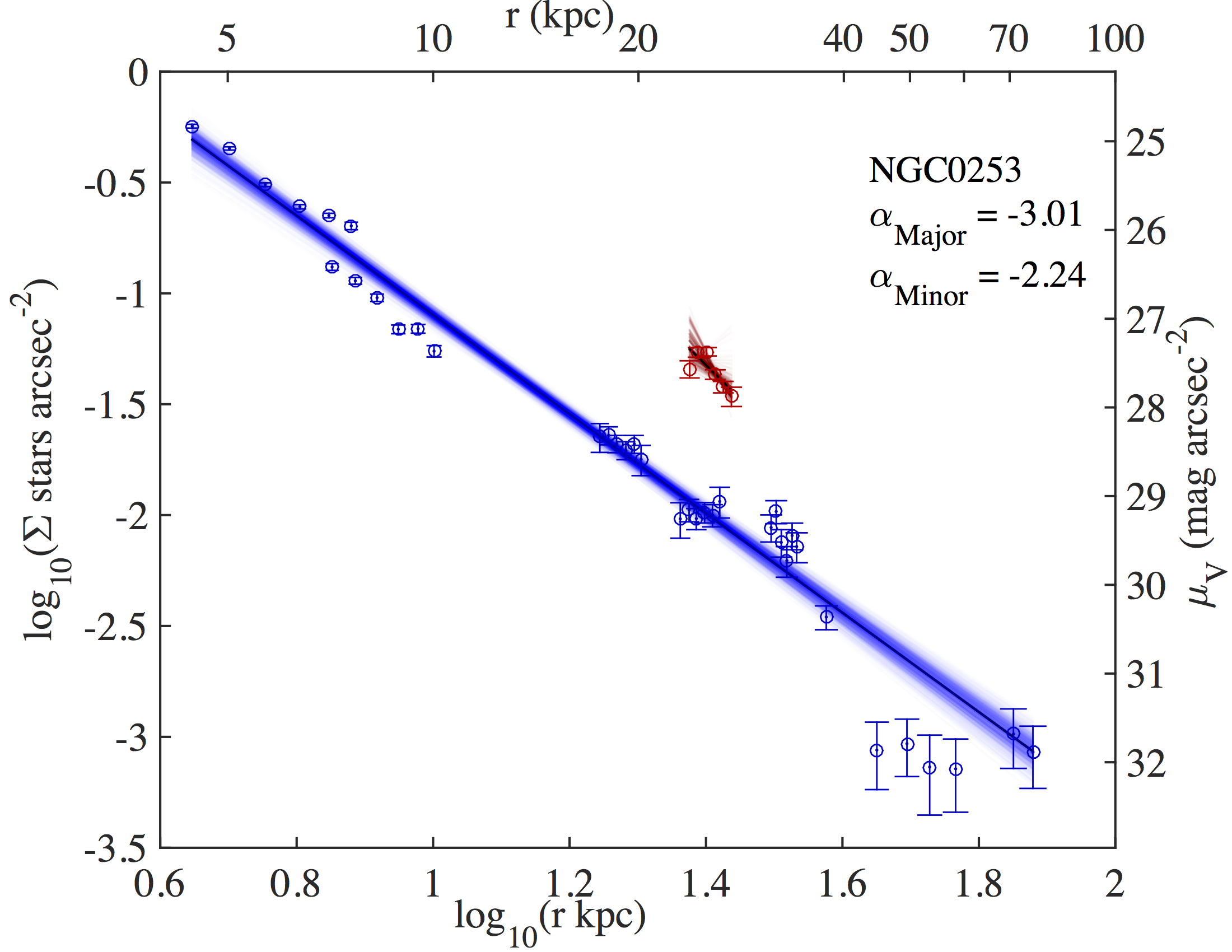}
	\caption{Stellar density profile for NGC 253's halo along the minor (blue) and major (red) axes. The line resulting from a maximum likelihood fit is displayed with its corresponding slope representing a best fit power-law for the halo. The translucent lines are the fits resulting from bootstrapping the data.}
    
    \label{NGC0253Profile}
\end{figure}

\begin{figure}\centering
	\includegraphics[width=80mm]{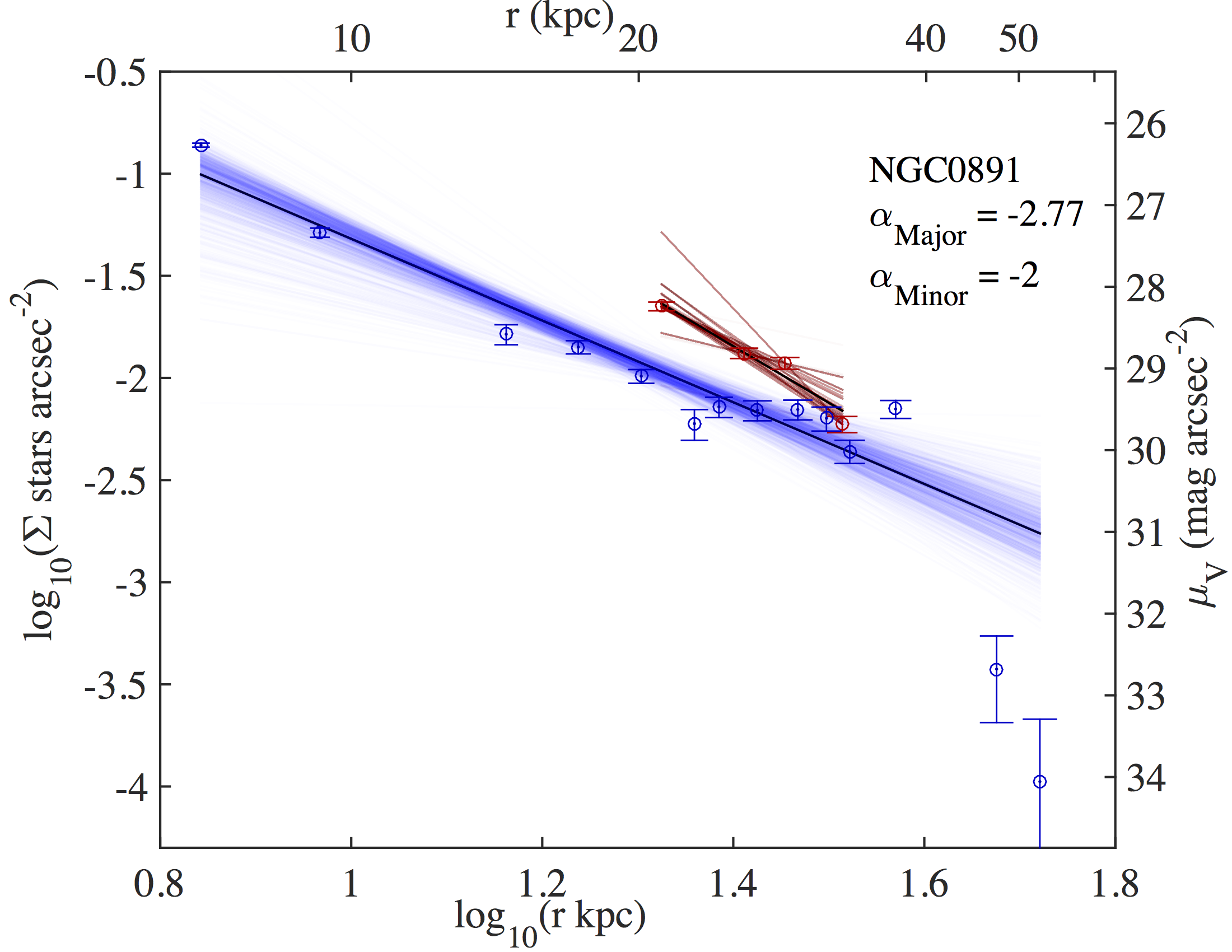}
	\caption{Stellar density profile for NGC 891's halo, for a general description, see Fig. \ref{NGC0253Profile}.}
    \label{NGC0891Profile}
\end{figure}

\begin{figure}\centering
	\includegraphics[width=80mm]{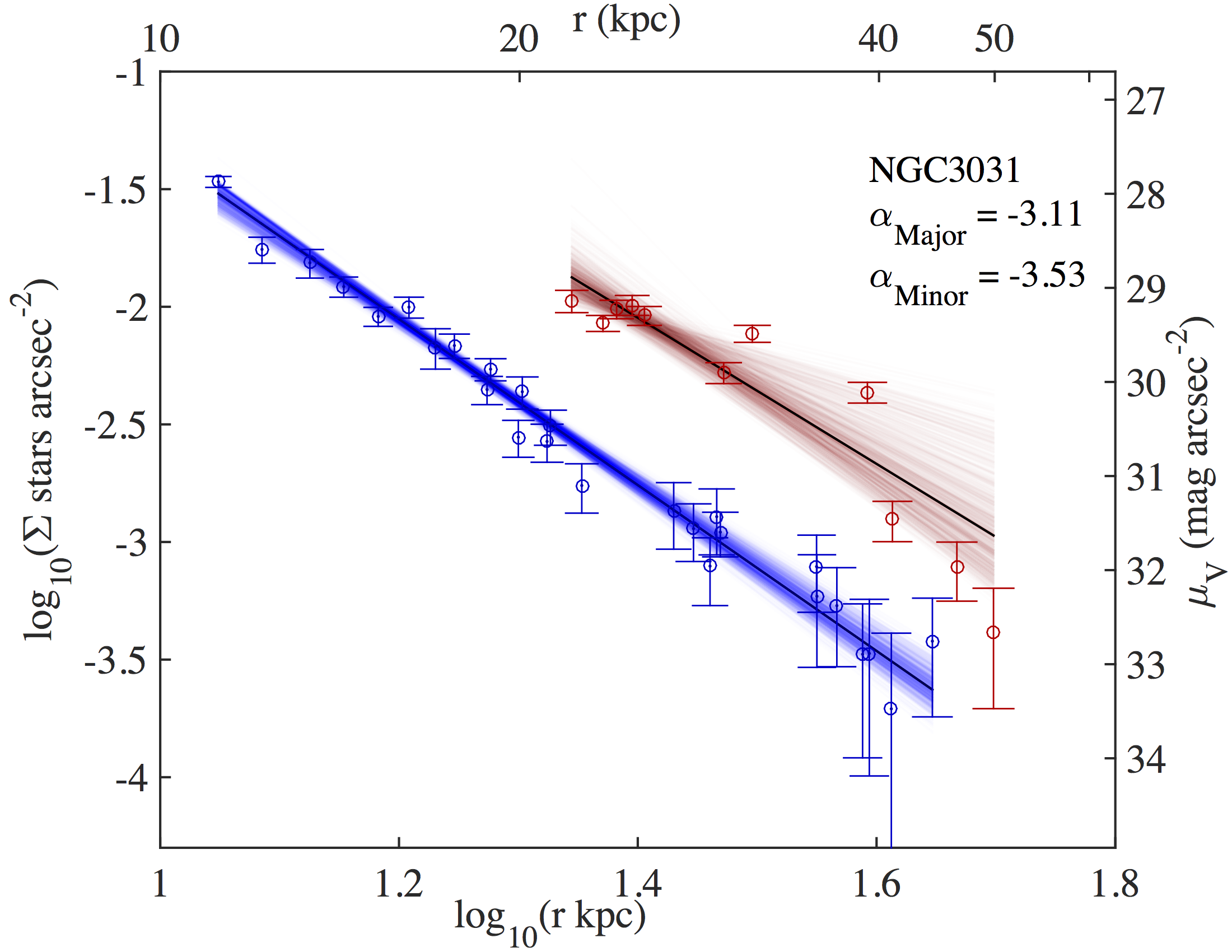}
	\caption{Stellar density profile for NGC 3031's halo, for a general description, see Fig. \ref{NGC0253Profile}. Part of the stellar halo of M82 can be seen in the major axis profile between $\sim$ 25 and 40 kpc as a significant overdensity.}
    \label{NGC3031Profile}
\end{figure}

\begin{figure}\centering
	\includegraphics[width=80mm]{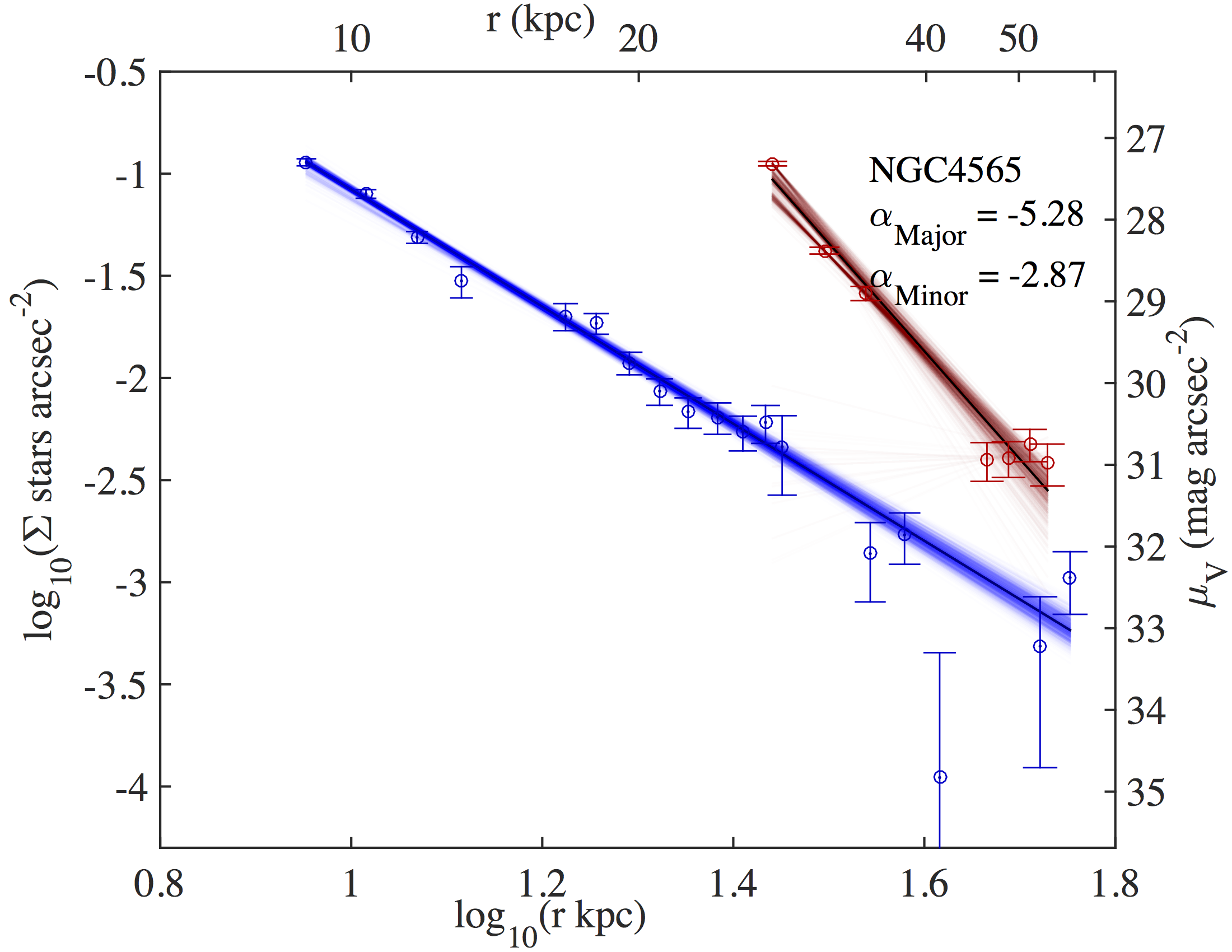}
	\caption{Stellar density profile for NGC 4565's halo, for a general description, see Fig. \ref{NGC0253Profile}.}
    \label{NGC4565Profile}
\end{figure}

\begin{figure}\centering
	\includegraphics[width=80mm]{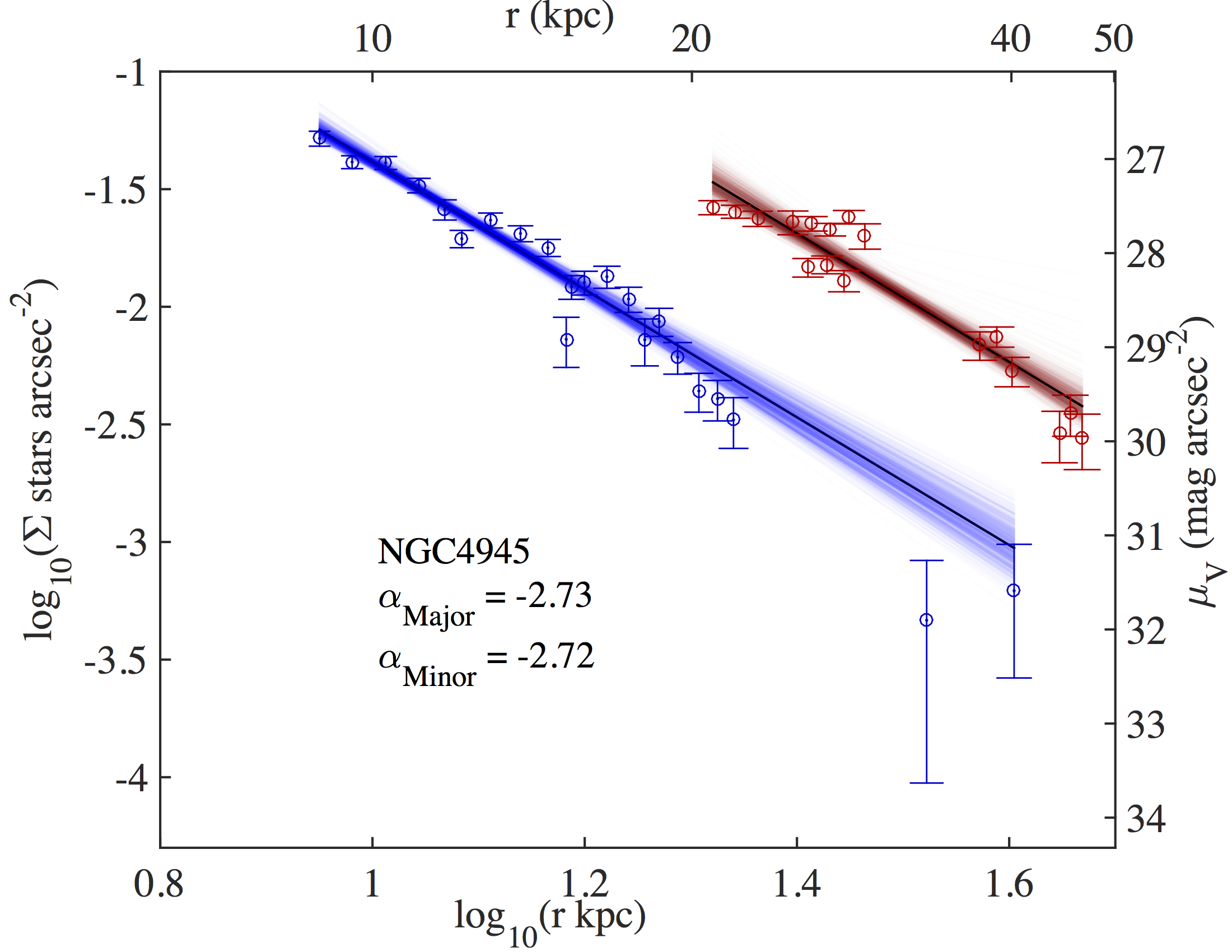}
	\caption{Stellar density profile for NGC 4945's halo, for a general description, see Fig. \ref{NGC0253Profile}
    }
    \label{NGC4945Profile}
\end{figure}

\begin{figure}\centering
	\includegraphics[width=80mm]{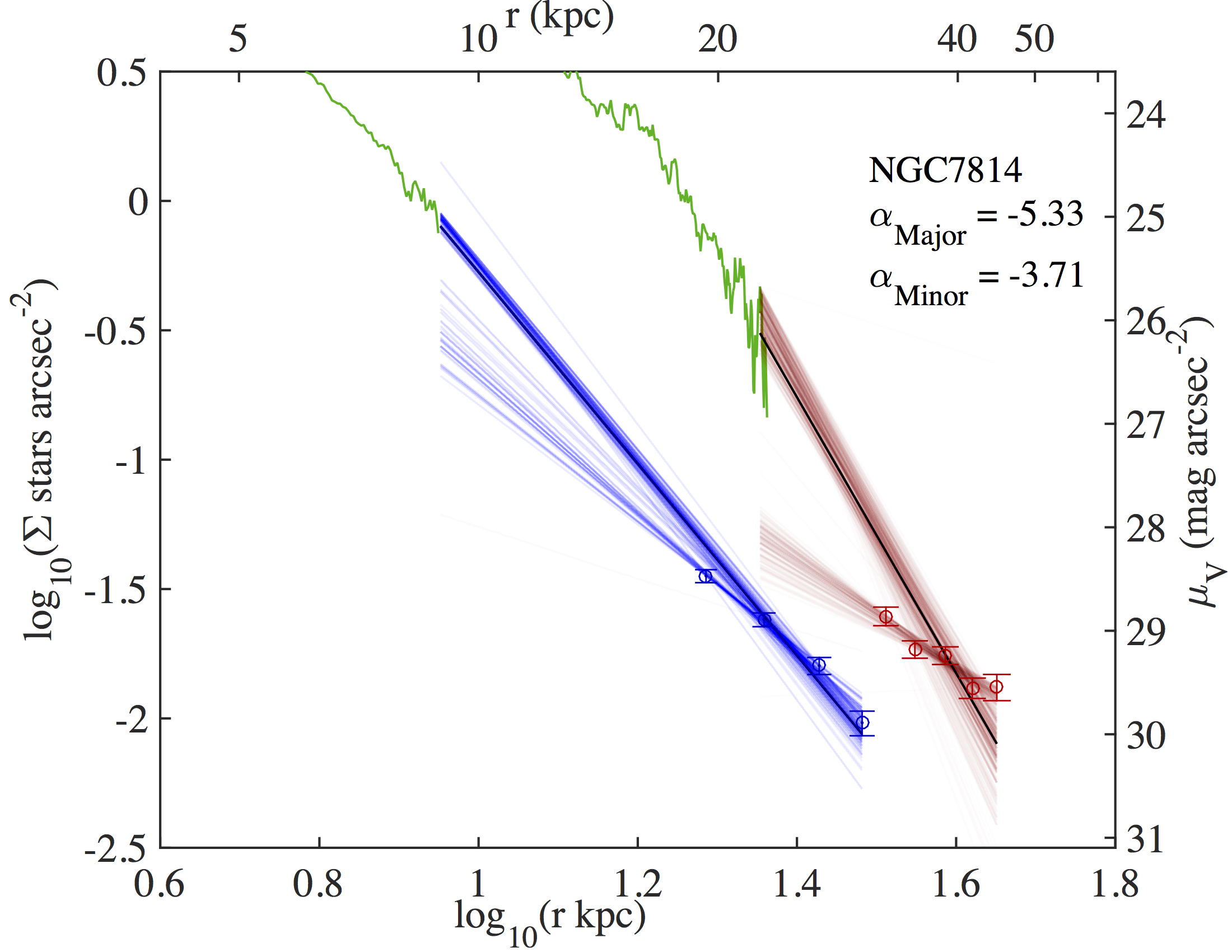}
	\caption{Stellar density profile for NGC 7814's halo, for a general description, see Fig. \ref{NGC0253Profile}. The green lines show the S4G integrated surface brightness profiles along the major and minor axes, converted into the equivalent star counts using isochrones.}
    \label{NGC7814Profile}
\end{figure}

Figs.~\ref{NGC0253Profile} to \ref{NGC7814Profile} show the stellar density profiles of each galaxy along their major (red symbols) and minor (blue symbols) axes. Given that many of the profiles appear to behave approximately as a power law with substantial scatter around that profile, we maximum-likelihood fit a three parameter power law model to each of the major and minor axis datasets for each galaxy. The fit is weighted based on the uncertainties in radial bin densities. We assume that the area density of RGB stars at a given projected radius $r$ can be drawn from the following distribution: 
\begin{equation}
P(\log_{10}\Sigma(r))=\frac{1}{\sqrt{2\pi\sigma}}e^{-\frac{(\log_{10}\Sigma(r)-\log_{10}\Sigma^\prime(r))^2}{2\sigma^2}},
\end{equation} 
where the expectation for the RGB star area density at that radius $\Sigma^{\prime}(r)$ is given by:
\begin{equation}
\log_{10}\Sigma^{\prime}(r)=\log_{10} \Sigma_{0}(r_{0})-\alpha\times \log_{10}(r/r_{0}),	
\end{equation} 
where $\Sigma_0$ is the density at the characteristic radius $\log r_0 = \left< \log r \right>$, $\alpha$ is the power law slope and $\sigma$ is the RMS (Table \ref{Table_Slopes_Masses}) of the data points around the expectation. The best fits are shown as the solid lines in Figs.\ \ref{NGC0253Profile} to \ref{NGC7814Profile}. Uncertainties were calculated by bootstrapping individual stellar density measurements, and the resulting boostrapped fits are shown in Figs.~\ref{NGC0253Profile} to \ref{NGC7814Profile} using translucent red or blue lines. The parameters for these power law fits are given in Table \ref{Table_Slopes_Masses}, and the star count values and isochrone-derived factors that we use to turn star counts into equivalent $V$-band surface brightness are given in Tables \ref{tab:datapoints} and \ref{tab:fluxstar} respectively.  

We find that the stellar density profiles of our sample of six roughly Milky Way mass galaxies decline steeply, broadly characterised by a range of power law functions $\Sigma \propto r^{-\alpha}$ where $1.7<\alpha<5.3$. For most galaxies there is substantial scatter around a single power law that is not well-described by measurement uncertainties alone, which is parametrised in this very simple model of a Gaussian scatter around the power law fit of up to $\sim 0.15$\,dex. This scatter appears to be systematic in nature, with coherent bumps and wiggles in the profiles, indicative of stellar halo substructure in the target galaxies. There is diversity in the recovered power law slopes in excess of the measurement uncertainties (the dispersion in slopes is substantially larger than the combined error for the slopes), indicative of real diversity in the stellar halo properties of these six roughly Milky Way sized galaxies. It is clear that the choice of a power law profile is an important over-simplification: a multi-part profile would be a substantially better fit for at least the minor axis profiles of NGC\,253 and NGC\,891 and possibly NGC\,4945, where the profiles appear to change slope at radii around $\sim$30 kpc.

The prominence of coherent brightness profile fluctuations (associated with recognised large-scale substructure in many galaxies; e.g., NGC 253 and NGC 891; see \S \ref{indiv}) acts to emphasise the importance of substructure in the study of stellar haloes. {\it All} of our observed and inferred characteristics --- the surface brightness profiles, mass, power law slope, intrinsic scatter, estimated axis ratio and stellar populations --- are influenced by these substructures. The properties of the stellar halo are best thought of as measurements of the `aggregate' stellar halo. The issue is whether one's survey has `fairly' sampled the different lines of sight to converge towards a `representative' measurement of stellar halo properties. Cognizant that this issue cannot be quantitatively settled without deep panoramic measurements for a large sample of galaxies (e.g., a survey like GHOSTS but with tens to hundreds of times more survey area), we provisionally estimate the magnitude of such effects using simulations in \S \ref{BJ_Test}. 

\subsection{Stellar Halo Axis Ratios}

Given the sparse sampling of GHOSTS along two principal axes, we have a relatively limited ability to estimate projected axis ratio. Given that the major axis profiles typically sample substantially less dynamic range in radius (from $\sim 20$\,kpc to roughly $\sim 40$\,kpc) than the minor axis profiles, we estimate axis ratio by comparing the minor and major axis density profiles at a characteristic radius of 25\,kpc. 

\begin{figure}\centering
	\includegraphics[width=80mm]{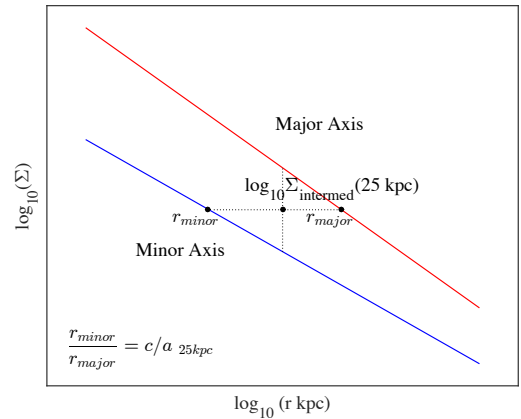}
	\caption{An illustration of the procedure for estimating the projected axis ratio $c/a$ at $\sim 25$\,kpc, $c/a_{\rm 25\,kpc}$. The average $\log_{10}\Sigma$ value at 25\,kpc is determined, and we estimate $c/a_{\rm 25\,kpc}$ to be $r_{minor}/r_{major}$. }
    \label{AxisRatioPlot}
\end{figure}

This `indicative' projected axis ratio $c/a_{\rm 25\,kpc}$ is determined using the power-law fits obtained above, as described in Fig.\ \ref{AxisRatioPlot}. Given the interpolated (slightly extrapolated in the case of NGC 4565's major axis) major and minor axis densities at 25kpc, the mean of the values of $\log_{10} \Sigma (25\,{\rm kpc})$ on the major and minor axis is calculated, $\log_{10} \Sigma_{\rm intermed} (25\,{\rm kpc})$. For each axis, the radius at which the interpolated (extrapolated only for NGC 253's major axis) densities reach $\log_{10} \Sigma_{\rm intermed} (25\,{\rm kpc})$ is recorded ($r_{minor}$ and $r_{major}$, as shown in Fig.\ \ref{AxisRatioPlot}), and we adopt $c/a_{\rm 25\,kpc} = r_{minor}/r_{major}$ as our best estimate of $c/a$. Formal uncertainties in $c/a_{\rm 25\,kpc}$ are calculated in concert with the power law fits to each axis, and are typically small ($<0.05$ in axis ratio), even when small extrapolations were necessary to estimate the value. In practice, there is considerable uncertainty in translating $c/a_{\rm 25\,kpc}$ into $c/a$, particularly in cases where the power law profiles of the major and minor axes differ considerably, indicating a radially-varying $c/a$. {We explore sources of systematic uncertainty in $c/a$ values using stellar halo models (\S \ref{BJ_Test}), finding a typical systematic uncertainty of $\Delta c/a \sim 0.1$, except in one case (out of 11) where there is a large misalignment between the model galaxy's principal axes and the stellar halo's principal axes. We adopt a systematic uncertainty of $\Delta c/a \sim 0.1$ in what follows.} The stellar halo axis ratio estimates $c/a_{\rm 25\,kpc}$ for the GHOSTS MW-mass galaxies range from $c/a\sim$0.4 to $\sim$0.75 (Table \ref{Table_Slopes_Masses}). 

\subsection{Stellar Halo Masses and Surface Brightnesses}

We determine the stellar halo mass $M_{10-40}$ between minor axis equivalent radii of $10-40$\,kpc, corresponding to $(10-40) [c/a_{\rm 25\,kpc}]^{-1}$ kpc along the major axis, using numerical integration. 
When determining the mass estimates, the choice of lower bound is particularly significant considering the divergent nature of a power-law fit. We chose 10 kpc as the inner bound since this is the closest galactocentric distance along the minor axis for which there is minimal to no disc contamination for the less highly-inclined galaxies, such as NGC 3031. The choice of outer bound has a relatively small effect; little mass lies outside 40\,kpc for the halo profiles characteristic of GHOSTS galaxies. We first integrate the minor axis power-law profile over the area of the halo within 10 kpc to 40 kpc, using elliptical annuli with a constant axis ratio of $c/a_{\rm 25\,kpc}$ to obtain the number of RGB stars within that area $N_{\rm RGB,10-40}$. We use stellar halo models in Section \ref{BJ_Test} to calibrate this measurement (which can be carried out equally well on our data and with models) and estimate how $N_{\rm RGB,10-40}$ and $M_{10-40}$ may be expected to compare to total stellar halo mass. 

We then use stellar evolution models to estimate the amount of mass and light represented by each detected RGB star. Our halo CMDs appear broadly consistent with old metal-poor populations; accordingly, we choose to adopt a fiducial 10 Gyr old Padova isochrone \citep{Bressan12, Chen14, Tang14} with a metallicity Z=0.0016 ($[\rm{Fe/H}] = -1.2$ dex) --- similar to the average metallicity for our dataset --- to represent the bulk of the halo population. We adopted a \citet{Chabrier03} stellar initial mass function (IMF). A well-populated model CMD was constructed, and the number of RGB stars in the selection region of the CMD (see red box in Fig.
\ref{CMDs}) per unit initial stellar mass and $V$-band luminosity is calculated for each galaxy. The right-hand axis on Figs.\ \ref{NGC0253Profile}-\ref{NGC7814Profile} shows the $\mu_{V}$ profile in units of $V-$mag arcsec$^{-2}$. 

Scaling of star counts to total surface brightness using stellar population models is a common technique \citep[e.g.,][]{Ibata14}. Nonetheless, it is useful to cross-validate our inferred surface brightness profiles with previously-published values. Such cross-validation is challenging owing to the difficulty in finding systems with low enough surface brightness for the resolved stellar populations to remain uncrowded while remaining well-measured in  integrated light ($V$-band surface brightnesses of $\sim 27$\,mag\,arcsec$^{-2}$). In addition, we wish to target metal-poor regions, as our star counts focus on metal-poor stars. 

We can compare our measurements of isochrone-scaled star counts with integrated surface brightness estimates for three systems in the GHOSTS sample: NGC 253, NGC 891, and NGC 4565. We compare our inferred $V$-band major axis surface brightness profile for NGC 253 with the $J$-band surface brightness profile of \citet[from star counts scaled to $J$-band brightness where their profiles overlapped]{Greggio14}, assuming $V-J \sim 1.7$ for a [Fe/H]$\sim -1$, 10Gyr old stellar population following \citet{Bruzual03}, finding agreement within $\Delta \mu \sim 0.1$ mag\,arcsec$^{-2}$.  We compare our inferred $V$-band minor axis surface brightness profile for NGC 891 at $6-9$\,kpc with the $R$-band brightness profile of \cite{Miller96} converted to $V$-band assuming a [Fe/H]$\sim -1$, 10Gyr old stellar population with $V-R \sim 0.52$ following \citet{Bruzual03}, finding agreement within $\Delta \mu \sim 0.3$ mag\,arcsec$^{-2}$. 
Turning to NGC 4565, a $<$10 kpc extrapolation of our inferred minor axis $V$-band brightness agrees within $\Delta \mu \sim 0.2$ mag\,arcsec$^{-2}$ with the $V$-band surface brightness of 27\,mag\,arcsec$^{-2}$ at 8\,kpc minor axis distance from \cite{NJ97}.
We conclude that our surface brightness measurements appear to be accurate, with no sign of a systematic offset at the 0.3\,mag\,arcsec$^{-2}$ level. 

The isochrones give estimates of initially-formed stellar mass, which must be corrected to present-day mass by accounting for stellar mass loss by multiplying the initially formed mass by 0.56 (following \citealp{Bruzual03}). The present-day stellar halo mass $M_{10-40}$ is then calculated by dividing the total number of detected RGB stars between minor axis equivalent radii of $10-40$\,kpc $N_{\rm RGB,10-40}$ by the number of RGB stars per unit present day stellar mass. Our resulting stellar halo mass estimates $M_{10-40}$ are presented in Table \ref{Table_Slopes_Masses}. We note that the random uncertainties (determined from bootstrapping) presented in Table \ref{Table_Slopes_Masses} do not include a contribution from systematic uncertainty about the halo stellar populations or isochrone uncertainties; we varied ages and metallicities by $\pm$\,30\% in age and a factor of three in metallicity, and this changes the final masses by $\pm$30\% or less. These are included in the systematic error budget in Table \ref{Table_Slopes_Masses}.

We also indicate in Table \ref{Table_Slopes_Masses} the total stellar mass of each galaxy, estimated using $K$-band luminosities in concert with a $K$-band mass to light ratio of $M/L=0.6$, typical of massive spiral galaxies, following \cite{BelldeJong01} using a universally-applicable \cite{Chabrier03} stellar IMF.  Luminosities were calculated using $K$-band total magnitudes from \cite{Jarrett03}, in conjunction with the distances presented in Table \ref{Table_Slopes_Masses}. Such masses carry at least 30\% uncertainties, and potentially suffer from larger systematic error if assumptions underlying their calculation are incorrect, e.g., if the stellar IMF varies from galaxy to galaxy. Despite these uncertainties, these masses are useful in order to build intuition about how these galaxies compare to larger samples of galaxies, e.g., from the SDSS (e.g., \citealp{Kauffmann03}) that have stellar mass estimates but lack accurate measures of rotation velocity.  

\section{How generalisable are our inferences from the data? Generating intuition through analysis of stellar halo models}
\label{BJ_Test}

Before examining the results for individual galaxies, inter-comparing them, and comparing our observations with theoretical models, it is important to generate intuition about how our results might generalise to the bulk properties of a realistically structured stellar halo. As articulated earlier, the key concern is the degree of systematic error caused by sparse sampling stellar halo structure in a highly structured aggregate halo; a secondary concern is the influence of stellar population variations in the stellar halo on our inferences. 

In the absence of panoramic imaging as deep or deeper than our data (e.g., future wide-area surveys with WFIRST and LSST), it is necessary to use simulations to explore this issue. While any simulation could be used in principle, we choose to analyze the 11 halo realisations from the \cite{BJ05} simulations\footnote{The stellar halo models are available at \href{url}{http://user.astro.columbia.edu/~kvj/halos/}.}. These stellar halo models are built through the disruption and accretion of satellite galaxies in a cosmological context. Star particles in sub-haloes were generated using high-resolution N-body simulations and painted on to dark matter particles such that their luminosity function follows a King profile. A cosmologically motivated semi-analytic galaxy formation model was used to assign stellar properties to the painted particles (see also \citealp{Robertson05}; \citealp{Font06a}).  We converted the star particles into RGB stars and generated projected RGB maps of stars as explained in \citet{Mon13}. For these haloes, we emulated ACS observations by choosing square sections of 202 arcseconds on a side along the major and minor axes. The different galaxy distances and colour-magnitude cuts that correspond to each of the six massive GHOSTS galaxies were used to examine the models. This allows us to determine how representative our data is for each galaxy. 

We choose to analyze 10 ACS-like fields per galaxy, 5 on the minor axis and 5 on the major axis. While clearly the number of pointings per galaxy varies from case to case (see Fig.\ \ref{fields}), this is close to the average number of independent pointings per galaxy. The simulated ACS-like fields were treated identically to the real ACS observations. A best-fit power-law was calculated and integrated over an ellipse between 10 and 40 kpc using an axis ratio derived at a `indicative' radius of 25 kpc, and the stellar mass of the models was found using the same process that was applied to the data. We compared the results from these simulated ACS observations to the true values for each model for the power-law slopes, axis ratio, and stellar halo mass as described below. 

\begin{figure*}\centering
	\includegraphics[width=150mm]{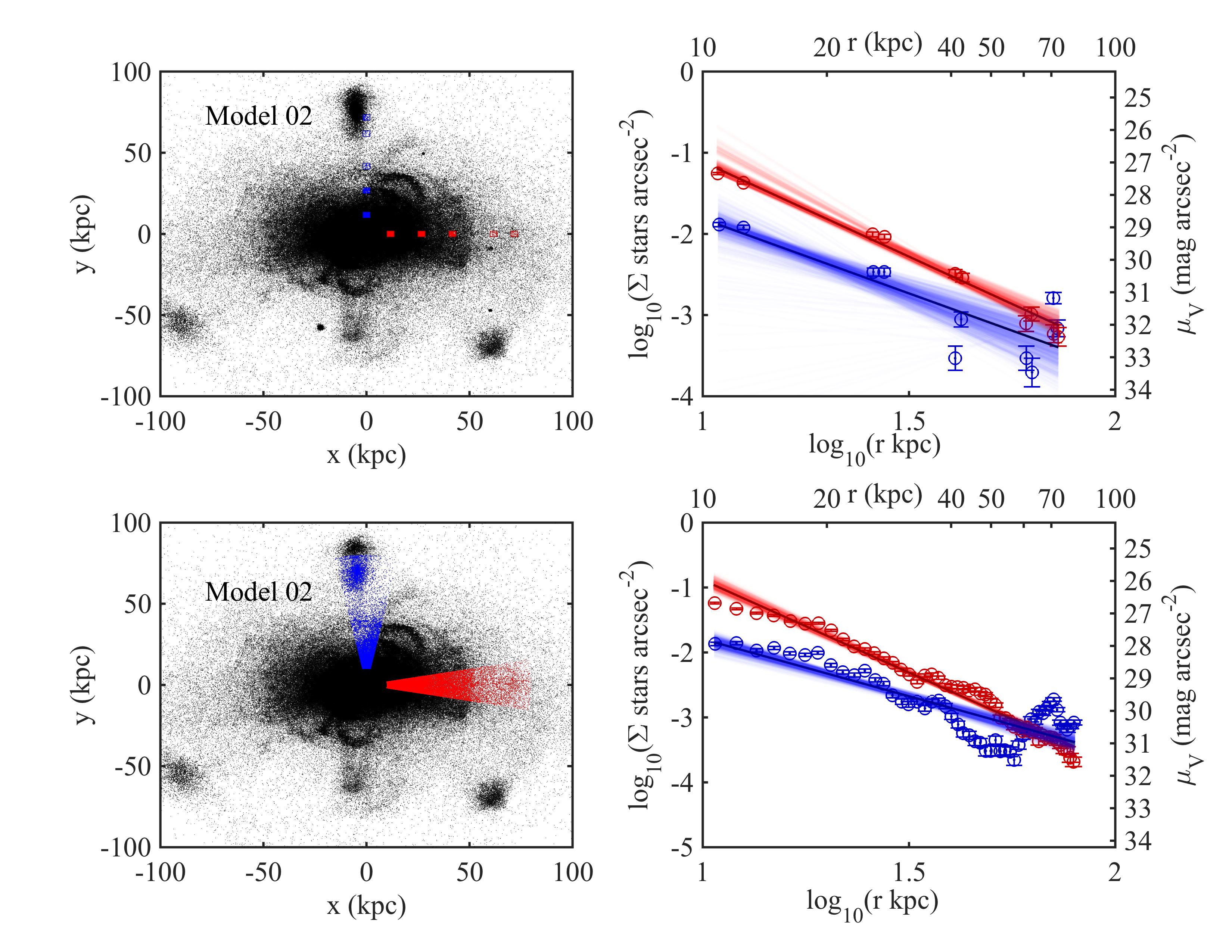}
	\caption{
    \textbf{Top:} Results from Bullock \& Johnston Halo 02 inferred using sparse sampling mimicking that of the data. ACS-like fields were placed along the minor and major axes as illustrated in the left panel, giving the stellar density profiles shown in the right panel.
The cuts in space and colour-magnitude were the same as those used in NGC 253.
    \textbf{Bottom:} Stellar density profile (right panel) and 2-D map of RGB stars (left panel) for Bullock \& Johnston Halo 02. The resulting density profiles along wedges on the major and minor axes are obtained from colour-magnitude cuts as those used in NGC 253. 
    The slopes for the major and minor axes as well as the masses calculated using the two different methods were in agreement within 10\%.}
    \label{ModelStars_Blades}
    \label{ModelStars_ACS}
\end{figure*}

In order to find the true power-law slope for the stellar density profile of the model, we selected stars within wedges of 1/8 radian half-width around the major and minor axes, between 10 and 80 kpc from the centre, as illustrated in the bottom-left panel of Figure \ref{ModelStars_Blades}. Each of these regions was divided into 50 radial sections and we constructed projected stellar density profiles of the modeled RGB stars on the minor and major axis. An example of the wedge density profiles for Bullock \& Johnston Halo 02 can be seen in the bottom-right panel of Figure \ref{ModelStars_Blades}.
The resulting power-law slopes that best fit the profiles were taken to be the ``true'' values in order to measure the accuracy of the simulated ACS observations. The top panel of Figure ~\ref{ModelStars_ACS} shows the ACS-like fields corresponding to the same Bullock \& Johnston Halo 02 model as well as the resulting density profiles. 
Comparing the results obtained using these two methods, we find that our sparse sampling method produces power law slope estimates accurate to about $\pm 0.2$.

To find the `true' axis ratio of the models at 25 kpc, we fit the RGB stars distribution of the models using an iterative method. We select RGB stars within an elliptical annulus with a geometric mean distance of 25\,kpc using an initial guess for axis ratio and assuming alignment between the major axis and the long axis of the initial ellipse. The second moment tensor of the distribution was calculated, giving improved estimates of axis ratios and position angle. This process was repeated until it converged to within 0.001 in axis ratio. We find the axis ratio we calculate based on sparse-sampled HST fields is accurate to within $\Delta c/a \sim 0.1$, except in one case where there is a large amount of substructure (1 case out of 11) where our method recovered $c/a \sim 1$ for a halo with actual $c/a_{\rm intrinsic} \sim 0.5$ owing to a misalignment between the actual position angle of the halo and the major axis of the galaxy. Given the sparse survey strategy that we adopted (constrained by the amount of available telescope time), it is difficult to guard against position angle differences between the halo and principal axes of the main body of the galaxy; given that this happens at the $1/11$ level in simulations, we expect the bulk of our axis ratios to be accurate to $\Delta c/a \sim 0.1$. 

These models also offer an important end-to-end test of our survey strategy's ability to infer reliable stellar halo masses. For a range of distances corresponding to our sample galaxies, we choose colour-magnitude selections appropriate to each galaxy and calculate the mass between minor axis equivalent radii of 10-40\,kpc using the method described above (using the minor axis profile and the indicative axis ratio at $\sim 25$\,kpc). In concert, we calculate the true mass between 10-40\,kpc in an elliptical annulus with the correct position angle and ellipticity (the `true' 10-40\,kpc mass) and the total stellar halo mass. Our observational and analysis techniques give estimates of $M_{10-40}$ which are $97\pm 22$\% of the `true' 10-40\,kpc mass; our estimates of $M_{10-40}$ correspond to $32\pm 10$\% of the total RGB stars for model stellar haloes from \citet{BJ05}. 

M81 presents a unique case as it has an inclination of 60$^{\circ}$. We rotate the models to simulate its orientation and find that the power-law slopes vary by typically less than 0.2 in power law slope, the masses by 10$\%$ or less, and the axis ratios increase typically by 0.2 compared to a perfectly edge-on model. {Accordingly, we include an extra systematic uncertainty of $^{+0.0}_{-0.2}$ in $c/a$ for M81 in Table \ref{Table_Slopes_Masses}.}

We incorporate estimates of these systematic uncertainties in Table \ref{Table_Slopes_Masses}. 

\section{Notes on individual galaxies}
\label{indiv}

Table \ref{Table_Slopes_Masses} presents our estimates of the stellar halo properties --- power law slope, normalisation, intrinsic scatter around a power law profile, indicative axis ratio and mass between minor axis equivalent radii of 10 and 40kpc,  for each of the galaxies studied. In this section, we discuss our results for individual galaxies and compare our estimates of halo properties, determined using our strategy which obtains deep high-quality detections on relatively few sparse pointings, with other work typically derived from wide-field ground-based studies. {In what follows, we will often quote random and systematic uncertainties separately.}

\subsection{NGC 253}
The minor axis density profile for NGC 253 is well-measured out to more than 75\,kpc, following a power-law with slope $-2.24^{+0.07}_{-0.06}\pm0.2$ (random and systematic errors respectively) reasonably well out to $\sim 50$ kpc as can be seen in Fig.\ \ref{NGC0253Profile}. We note that this detection of stellar halo stars at $>75$\,kpc is somewhat remarkable --- only three galaxies, the Milky Way, M31 and Centaurus A \citep{Rejkuba14,Crnojevic16}
have halo stars detected to such radii.

There is significant scatter around the fitted power-law profile, with a best fit intrinsic RMS of $0.10\pm0.01\pm0.03$ dex (random and systematic errors respectively). These deviations are systematic, with coherent over-densities compared to the power law fit at $\sim 30$\,kpc, and coherent underdensities at $\sim 10$\,kpc and most notably outside $\sim 40$\,kpc, where the profile is significantly depressed compared to smaller radii and appears to become flat. Comparison with the single halo-dominated major axis field in the GHOSTS survey yields a rough estimate of c/a $\sim 0.55^{+0.04}_{-0.05}\pm0.1$ (random and systematic errors respectively) for the projected axis ratio, though the uncertainties may be larger owing to our use of only a single major axis field. 

There are two existing estimates of the power law slope and axis ratio of the stellar halo of NGC 253 from panoramic ground-based imaging: \cite{Greggio14} used VISTA wide-area near-infrared imaging to determine a slope of $\sim -1.6$ and an axis ratio $b/a \sim 0.4$, and \cite{Bailin11} measured a power-law slope of $-2.8\pm0.6$ and $b/a \sim 0.35$ using IMACS data for the southwest quadrant of NGC 253's stellar halo. Our power law slopes are intermediate to these estimates, and our axis ratio is rather larger than both of these estimates. These works, along with our own and that of \citet{Davidge10}, all show clear evidence of substantial substructure in NGC 253's stellar halo. Two significant over-densities have been reported: a prominent `shelf' in the southwestern quadrant of the inner part of NGC 253's halo \citep{Beck82,Davidge10,Bailin11,Greggio14}, and an overdensity along the northern minor axis at $\sim 30$\,kpc best visualised in Figs.\ 16 and 21 of \citet{Greggio14}. Our minor axis profile intersects the northern minor axis overdensity, and it is clearly visible in Fig.\ \ref{NGC0253Profile} as an overdensity at 30\,kpc, beyond which the star count profile drops precipitously. We interpret the significant differences in stellar halo parameters reported by our work, \citet{Greggio14} and \citet{Bailin11} to stem in large part from the prominent substructure in NGC 253's stellar halo (this was also emphasised by \citealp{Bailin11} and \citealp{Greggio14} as their main source of systematic uncertainty); such differences may indicate the level of variation expected from study to study owing to substructure in stellar haloes.

Only one estimate of stellar halo mass has been published to date: $2.5 \pm 1.5 \times 10^{9}$ $M_{\odot}$ outside of minor axis radius of 5\,kpc (4.5\% of the galaxy stellar mass) from \cite{Bailin11} using wide-area coverage of the southwestern quadrant of the inner parts of NGC 253's halo. Our halo mass estimate is $1.45^{+0.17}_{-0.10}\pm0.5\times 10^{9}$ $M_{\odot}$ (random and systematic errors respectively) between minor axis equivalent radii of 10-40\,kpc; recall in \S \ref{BJ_Test} we use models to suggest that this likely implies a 3 times larger total stellar halo mass, implying a total stellar halo mass of roughly $4.5\pm1.9\times 10^{9} M_{\odot}$ ($8\pm3$\% of the galaxy stellar mass). These estimates agree to within their uncertainties. 

\subsection{NGC 891}

As far as we are aware, our measurement is the first quantitative measurement of the stellar halo density profile, axis ratio and mass for NGC 891. In particular, the mass of the stellar halo between 10 and 40 projected minor axis equivalent kpc is $8.6^{+0.7}_{-0.5}\pm2.6 \times 10^8 M_{\odot}$ (random and systematic errors respectively). This corresponds to an estimated total stellar halo mass of $2.7\pm1.2 \times 10^9 M_{\odot}$, corresponding to $5\pm2$\% of NGC 891's total stellar mass. 
NGC 891 has been imaged using Subaru's SUPRIME-CAM \citep{Mouhcine10}, leading to the discovery of extensive stellar streams and a relatively dense `cocoon' of stars in the inner parts of NGC 891's stellar halo (their Fig.\ 1). We clearly detect the stream and cocoon (towards \citealp{Mouhcine10}'s positive Z direction) on the minor axis fields between 25 and 40\,kpc, where the density profile is close to flat. This overdensity, and relatively dramatic drop in density outside 40\,kpc, drive both a relatively uncertain minor axis power law slope ($-2.00^{+0.33}_{-0.23}\pm0.2$) and one of our largest values of intrinsic scatter ($0.13\pm0.05\pm0.03$ dex). One could arbitrarily fit the density profile with a double power law broken at $\sim$ 40\,kpc, in which case the best fit slopes are $\sim -2$ inside 40\,kpc and $\sim -7$ (but with huge uncertainty) outside 40\,kpc. We do not adopt the parameters of this fit in this work, nor do we show it on Fig.\ \ref{NGC0891Profile}; such a fit would be too specific to the particular density profile seen in Fig.\ \ref{NGC0891Profile} and would hinder fair comparison with other galaxies or with simulations (most of which use single power-law fits to broadly characterise the density distribution). 

\subsection{NGC 3031/M81}
\label{m81profdisc}

\begin{figure}\centering
	\includegraphics[width=75mm]{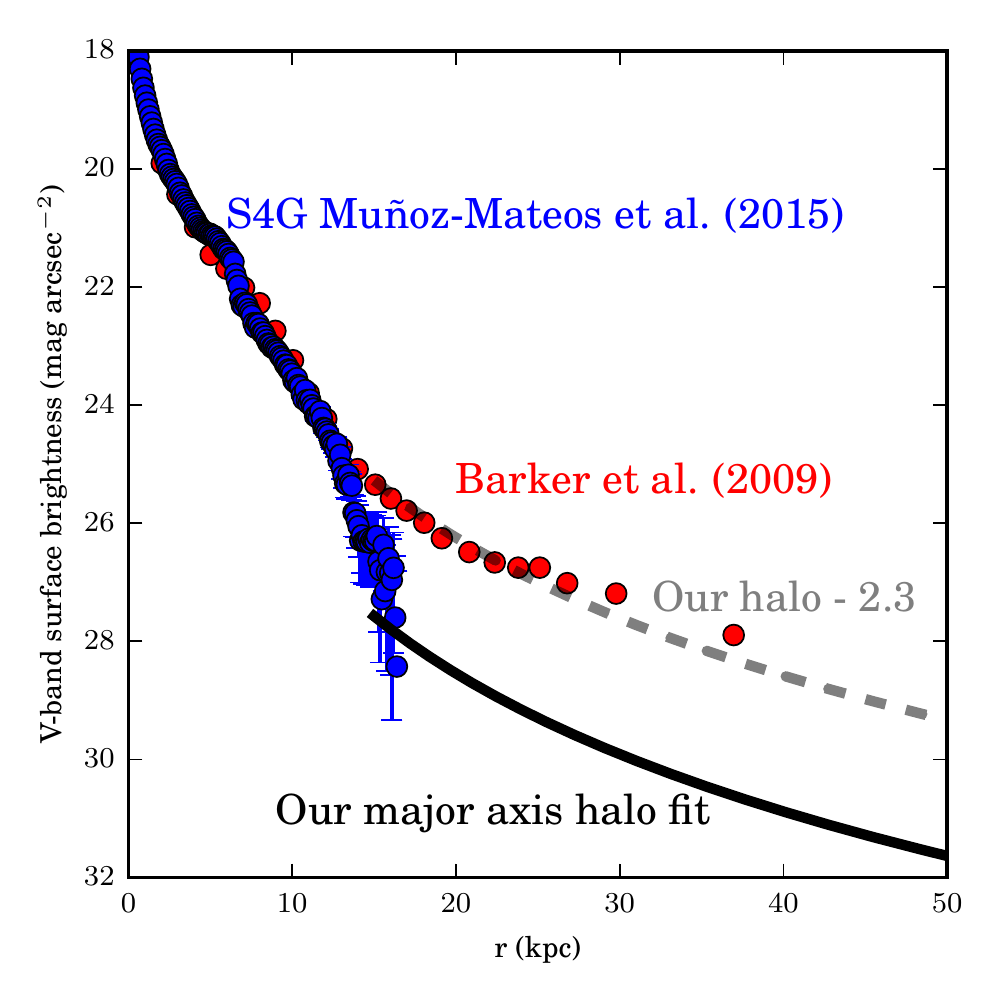}
	\caption{Major axis V-band surface brightness profile M81/NGC 3031. We show the 3.6$\micron$ surface brightness profile from S4G \citep{MM15} scaled to $V$-band by matching the inner parts of Barker et al.'s surface brightness profile, the final combined integrated light$+$star counts profile from \protect\citet{Barker09} in red, and our major axis halo fit in black, along with our fit rescaled 2.3\,mag\,arcsec$^{-2}$ brighter to approximately match the \protect\citet{Barker09} profile in their outer parts in dashed-gray line. It is clear that the \protect\citet{Barker09} star counts were erroneously calibrated to a surface brightness around $10\times$ higher than the isochrone-derived calibrations we use, and predict far more surface brightness than either we or \protect\cite{MM15} observe at deprojected major axis radii in excess of 14kpc.  }
    \label{M81_Barker}
\end{figure}

Power law fits over a dynamic range of a factor of 4 in radius for NGC 3031/M81 along the minor axis and a factor of nearly 2.5 in the major axis show that the metal-poor RGB stars show a steeply-declining roughly power law profile with slopes $-3.53^{+0.18}_{-0.15}\pm0.2$ and $-3.11^{+0.88}_{-0.48}\pm0.2$ respectively. The scatter is $0.03\pm0.02 \pm0.03$\,dex and $\sim 0.14^{+0.04}_{-0.06}\pm0.03$\,dex along the minor and major axis respectively, and axis ratio is {$0.61^{+0.03}_{-0.05}\pm0.1^{+0.0}_{-0.2}$, where the last error term accounts for the possible increase of projected axis ratio compared to the intrinsic axis ratio owing to M81's intermediate inclination}. This yields a stellar halo mass between minor axis equivalent radii of $10-40$\,kpc of $3.7^{+0.4}_{-0.2}\pm1.1\times 10^8 M_{\odot}$.  This corresponds to an estimated total stellar halo mass of $1.1\pm0.5\times 10^9 M_{\odot}$, corresponding to $2\pm0.9$\% of M81's total stellar mass. 

Many of our values appear to be in significant conflict with the only other estimates of the properties of M81's stellar halo from \citet{Barker09} using ground-based Suprime-Cam observations. While our axis ratio estimate of $\sim 0.6$ is in agreement with the axis ratio of $\sim 0.5$ 
{\it assumed} by \citet{Barker09} when analyzing the inner part of M81's stellar halo, our other measurements disagree with those of \citet{Barker09}. Our power-law slopes are $\sim -3.5$, whereas those of \citet{Barker09} are $\sim -2$, and most prominently, our estimated total stellar halo mass of $1.1\pm0.5\times 10^9 M_{\odot}$  (corresponding to $\sim 2$\% of M81's total stellar mass) appears to differ by almost an order of magnitude with their claim that M81's halo contains 10-15\% of the luminosity of M81.

We explore this discrepancy in depth in Fig.\ \ref{M81_Barker}, which shows the major axis V-band surface brightness profile M81/NGC 3031 from Fig.\ 17 of \protect\citet{Barker09} in red, and our major axis halo fit in black. These are clearly discrepant at the radii at which they overlap, but are not grossly different in shape, as evidenced by the dashed gray line, which shows our major axis profile fit offset by 2.3 magnitudes to approximately overlap with \citet{Barker09}\footnote{The difference in power law slope is visible by a `drift' in the best offset between the two datasets of about 0.5mag between 20 and 40kpc, in the sense that the brightness profile of \citet{Barker09} is flatter than ours.}. This brightness offset (coupled with minor differences in extrapolations to total stellar halo mass and luminosity) accounts for the difference between our and their halo luminosity estimates.

How is such a large difference in calibration possible? We attempt to shed light on this issue by comparing these brightness profiles with the 3.6$\micron$ surface brightness profile from S4G \citep{MM15} scaled to $V$-band by matching the inner parts of Barker et al.'s surface brightness profile. S4G \citep{Sheth10} is sensitive to relatively faint levels, and is much more immune to low surface brightness Galactic cirrus emission than optical light (clearly visible in Fig.\ 2 of \citealp{Barker09}). The S4G brightness profile --- well-measured out to $\sim 17$\,kpc --- clearly continues to decline with an exponential profile well outside of $\sim 12-14$\,kpc where \citet{Barker09} claim a transition in the integrated brightness profile to a shallower power law. As discussed in their Section 6 and shown in their Fig.\ 17, \citet{Barker09} use their 14-17\,kpc integrated light profile (which we believe to have been erroneously bright owing to Galactic foreground cirrus, and as evidenced by the S4G data) to calibrate their star counts. By enforcing that they overlapped, \citet{Barker09} calibrated their star counts to have a surface brightness nearly a factor of 10 brighter than those that they would have derived by doing artificial star tests plus calibration with isochrones.

It is interesting to ask what drives the difference in power law slope reported by \citet{Barker09} ($\Sigma \sim r^{-2}$) and our derived slopes along minor and major axes (both $\Sigma \sim r^{-3}$). Part of this difference could well be crowding in the ground-based data.   \protect\citet{Bailin11} show for NGC 253 (with a very similar distance to M81) that within radial distances of $\sim 20$\,kpc the ground-based data were very crowded, resulting in the detection of only a tiny fraction of the real RGB stars, particularly at brighter surface brightnesses, artificially flattening their star counts.\footnote{Incidentally, we also attribute the lack of a strong change in slope in the \citeauthor{Barker09} star count profile within $\sim 17$\,kpc, where S4G predicts a transition from an exponential to power law profile, to the effects of crowding.} Furthermore, real substructure may be responsible for some of the difference in power law slope: inspection of the major axis profile in Fig.\ \ref{NGC3031Profile} between major axis radii of 20\,kpc and 40\,kpc (the range covered by \citealp{Barker09}) shows a considerably flatter profile ($\Sigma \propto r^{-2}$ or somewhat shallower) than the profile between 20\,kpc and 50\,kpc ($\Sigma \propto r^{-3}$). Recently, \citet{Okamoto15} showed that there is an extensive fan of debris between M82 and M81 (we expect composed largely of material tidally liberated from M82). This is clearly detected along our major axis fields, and is prominent in much of the area probed by \citet{Barker09}. We propose that this drives the density profile derived by \citet{Barker09} towards $\Sigma \propto r^{-2}$. We claim that our measurement of a projected density profile $\Sigma \propto r^{-3}$ is somewhat more representative of M81's stellar halo not only because it is derived from the minor axis where there appears to be no material stripped off of M82 or NGC 3077 \citep{Okamoto15}, but also because it draws from a larger range of major axis radii, showing a return to a $\Sigma \propto r^{-3}$ profile outside of projected major axis radii of 40\,kpc. 

We conclude that the {\it shape} of the brightness profiles from our work and \citet{Barker09} are largely consistent, given the importance of both crowding and substructure on the Barker et al. result.  Our more reliable isochrone-based luminosity and stellar mass calibration differs strongly from Barker et al.'s, and with the benefit of deeper uncrowded HST data and S4G's deep integrated light profile we conclude that Barker et al.'s brightness calibration and luminosity estimate appears to be in error, owing to an unfortunate limitation in how their star counts were converted into an estimate of the $V$-band surface brightness.  

\subsection{NGC 4565}

As far as we are aware, this work, \cite{M16a}, and preliminary results from \cite{deJong09} are the first reported detection and characterisation of NGC\,4565's resolved stellar halo. NGC\,4565's stellar halo is detected out to 60\,kpc along the minor axis, and more than 50\,kpc along the major axis. NGC 4565's minor axis density profile has a power law slope of $-2.87^{+0.08}_{-0.07}\pm0.2$ (random and systematic uncertainties respectively), and while the fit prefers $\sim 0.11\pm0.03$\,dex of intrinsic scatter, owing to the magnitude of the error bars in the outer parts of NGC 4565 it is also consistent with having no intrinsic scatter around its minor axis profile. NGC 4565 has a substantially steeper major axis profile, with a power law slope of $-5.28^{+0.47}_{-0.45}\pm0.2$. If interpreted in terms of a halo with changing projected axis ratio, the axis ratio would vary from $c/a_{\rm 25\,kpc} \sim 0.44$ to $c/a_{\rm 40\,kpc} \sim 0.56$ (determined from the outermost points of the major axis profile, comparing them to the points of equal density along the minor axis at radii $\sim 30$\,kpc). 
The calculated stellar mass between 10 and 40 minor axis equivalent \,kpc is $7.2\pm0.3\pm2.2\times 10^8 M_{\odot}$, corresponding to a estimated total stellar halo mass of $2.2\pm0.9\times 10^9 M_{\odot}$ or $2.8\pm1.2$\% of the total stellar mass of NGC 4565. 

We note that Field 06 (the outermost minor axis field in NGC\,4565) has a significant overdensity (the high datapoint at minor axis radius $\sim 57$\,kpc), which we interpret to be a relatively thin stellar stream (also discussed in \citealp{M16a}). The width of the overdensity is $\sim 2-3$\,kpc. 

\subsection{NGC 4945}

As far as we are aware, this work (together with \citealp{M16a}) is the first reported detection and characterisation of NGC 4945's stellar halo. Owing to the substantial foreground contamination, we detect the minor axis to distances of only $\sim 40$\,kpc and the major axis to $\sim 45$\,kpc. The power-law slopes for the density profiles along the minor and major axes are consistent with each other at $-2.72\pm0.17\pm0.2$ and $-2.73\pm0.23\pm0.2$, respectively. The scatter around the minor axis profile is $0.05^{+0.01}_{-0.02}\pm0.03$\,dex, and major axis is $0.09^{+0.01}_{-0.02}\pm0.03$\,dex. Owing to the similarity of the density profiles along the major and minor axis, the axis ratio appears to be $c/a \sim 0.5\pm0.1$ with little radial dependence (although it is measured only out to minor axis equivalent radii of $\sim 22$\,kpc). The resulting halo stellar mass between 10 and 40 kpc is $1.11^{+0.07}_{-0.06}\pm0.33\times 10^9 M_{\odot}$, corresponding to a rough estimate of total stellar halo mass of $3.5\pm1.5\times 10^9 M_{\odot}$ which is roughly $9\pm4$\% of the total stellar mass of NGC 4945.

\subsection{NGC 7814}
\label{7814Discussion}

NGC 7814 is the most distant galaxy in our sample; as can be seen in Fig.\ \ref{CMDs}, the CMDs are relatively shallow and are limited in their colour coverage towards the red fainter than F814W$\sim 27.2$ by a relatively shallow F606W limit. Accordingly, we caution that our stellar counts may be somewhat less reliable than for our other galaxies, and may represent a lower limit to the true value. {Furthermore, the innermost fields in GHOSTS suffer from significant crowding. Reliable GHOSTS star counts exist only between $\sim 19$ and $\sim 35$ kpc along the minor axis and beyond $\sim 32$ kpc along the major axis. We supplemented the minor (major) axis profile with an equivalent star count value at $\sim$9 ($\sim$23) kpc derived from 3.6{\micron} imaging data from S4G, with a normalization derived using the same isochrone used to convert the RGB star counts into $V$-band surface brightness and stellar mass. } NGC 7814 has power slopes of $-3.71^{+0.99}_{-0.09}\pm0.2$ along its minor axis, and $-5.33^{+3.34}_{-0.57}\pm0.2$ along its major axis. The implied axis ratio is $0.59^{+0.14}_{-0.05}\pm0.1$. The implied stellar mass between 10 and 40kpc is $2.05^{+0.43}_{-0.26}\pm0.6\times 10^9 M_{\odot}$, or a estimated total stellar halo mass of $6.41^{+1.34}_{-0.81}\times 10^9 M_{\odot}$, corresponding to $\sim 14\pm6$\% of its total stellar mass. {This is the largest stellar halo in our sample}. NGC 7814 has a wide-format imaging from observations with small robotic telescopes \cite{Martinezdelgado10} and
does not show any signs of tidal streams \citep{Javanmardi16}. 

\section{Discussion and Model Comparisons}
\label{disc}

\begin{figure*}\centering	
	\includegraphics[width=170mm]{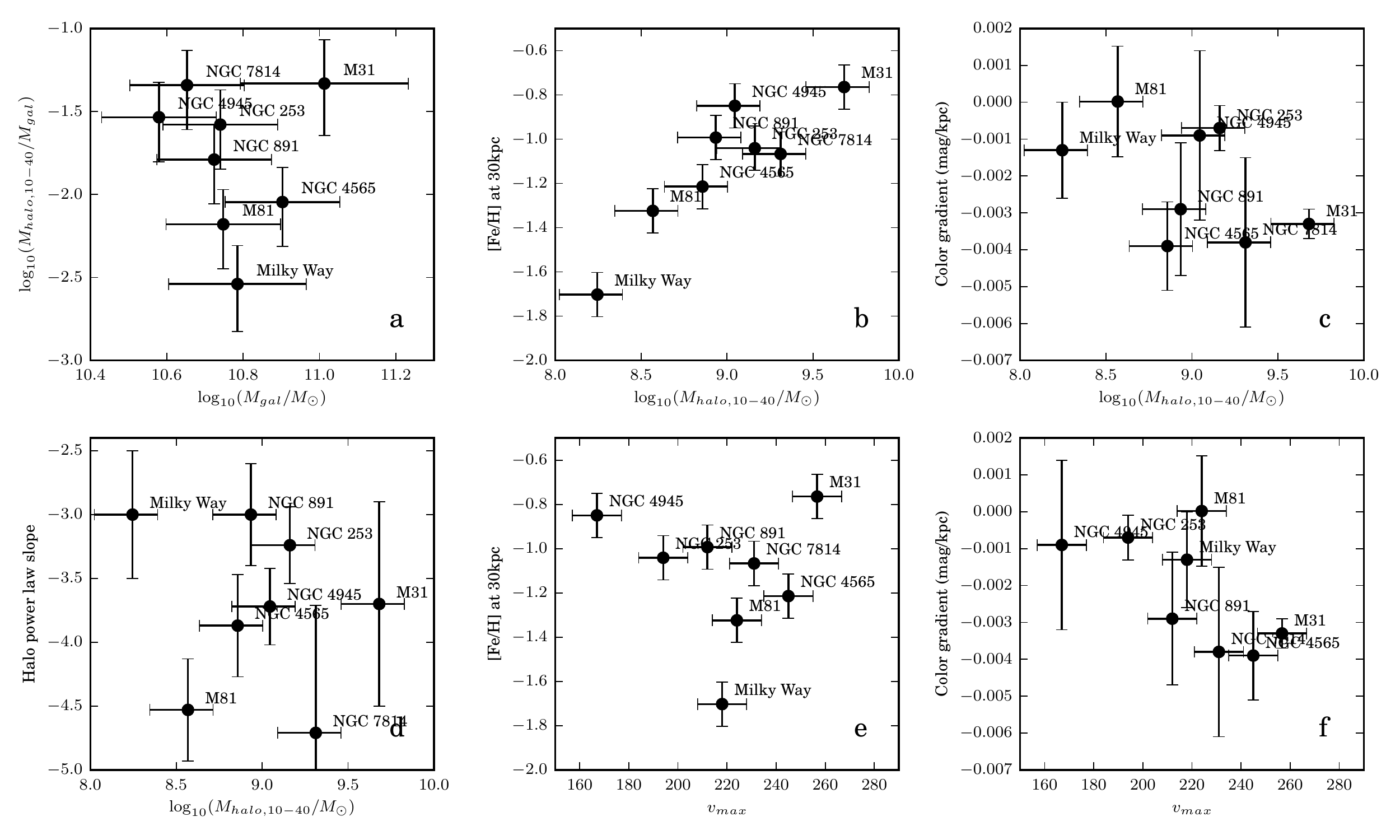}
	\caption{Panel $a$: ratio of stellar halo mass between 10-40\,kpc and total stellar mass, as a function of total stellar mass. Panel $b$: stellar halo metallicity at 30\,kpc as a function of stellar halo mass between 10-40\,kpc. Panel $c$: stellar halo colour gradient (a proxy for metallicity gradient) as a function of stellar halo mass between 10-40\,kpc. Panel $d$: inferred 3-D minor axis stellar halo density power law slope as a function of stellar halo mass between 10-40\,kpc. Panel $e$: stellar halo metallicity at 30\,kpc as a function of maximum rotation velocity. Panel $f$: stellar halo colour gradient (a proxy for metallicity gradient) as a function of maximum rotation velocity.}
	\label{correlations}
\end{figure*}

\subsection{Comparison of Stellar Halo Properties Between Galaxies} \label{sec:comp_obs}

With the inferred stellar halo masses, axis ratios and power law slopes in hand, along with an idea of the likely sources of systematic uncertainty, we turn to exploring correlations among the GHOSTS galaxies between these halo properties and their halo stellar populations 
\citep{M16a} as well as compare with the bulk properties of the Milky Way's and M31's stellar haloes.

We restrict our comparisons to quantities which are well-constrained by the data in hand. Accordingly, we choose to characterise the stellar halo mass using the mass between 10 and 40 minor axis equivalent kpc, $M_{10-40}$; minor axis power law slopes are measured over a similar range. We characterise the stellar halo metallicity by quoting a derived [Fe/H] value at 30 kpc along the minor axis following the observational calibration of [Fe/H] as a function of RGB colours for globular clusters \citep{Streich14} assuming $[\alpha/{\rm Fe}] =0.3$. Instead of presenting minor axis metallicity gradients, we choose to present minor axis RGB colour gradients per kpc; RGB colour is related in a highly non-linear way to metallicity, making RGB colour gradient more robust to possible future changes in RGB colour calibration than an inferred metallicity gradient. 

\begin{figure}\centering	
	\includegraphics[width=90mm]{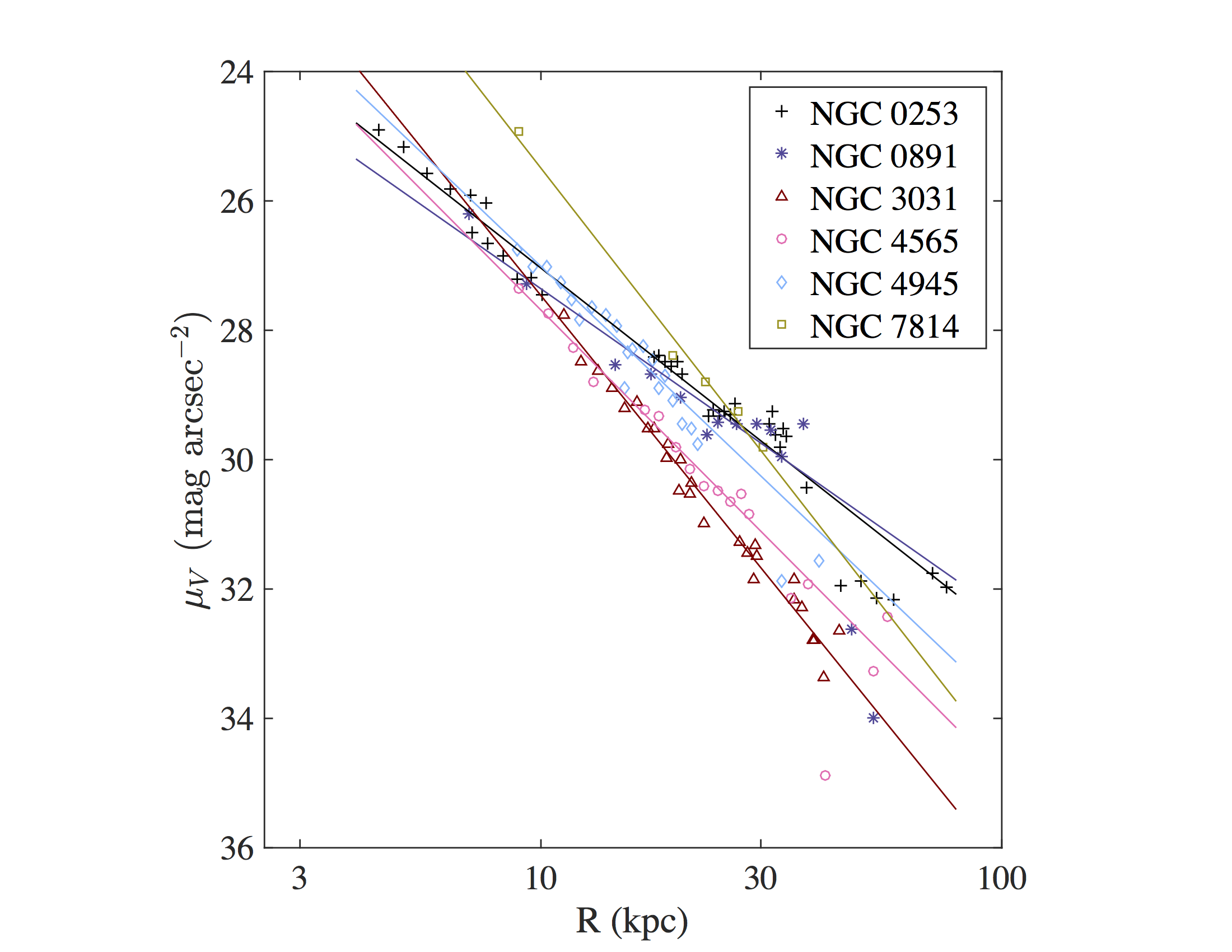}
	\caption{ Surface brightness profiles for the minor axes of GHOSTS massive disc galaxies. The plot shows the data  coloured for each galaxy, converted from star counts into $V$-band magnitude per arcsec$^2$ together with the power law best-fit obtained for the profiles.}
	\label{surfacebrightness}
\end{figure}

The properties of M31 and the Milky Way are compiled from a variety of sources. 
Stellar masses for the MW and M31 are assumed to be $6.1\pm1.1 \times 10^{10}M_{\odot}$ and $10.3\pm2.3 \times 10^{10}M_{\odot}$, taken from \cite{LiqNew15} and \cite{Sick15}, respectively. 
Rotation velocities for the MW and M31 are adopted from \cite{Bovy12}; $V_c = 218\pm6$\,km/s and HyperLEDA\footnote{\url{http://leda.univ-lyon1.fr/}; see also \cite{Makarov14}. } $V_{c,\rm{M31}} = 257\pm6$\,km/s, respectively.
The stellar halo mass for M31 outside 27 kpc is estimated to be $\sim 1.1\times 10^{10}M_{\odot}$ and has a 3-D (2-D) density slope of roughly $\sim -3.7$ ($-2.7$) \citep{Ibata14}. Extrapolation of the profile inside 27 kpc is obviously uncertain; we assume a total mass $M_{halo,{\rm M31}} = 1.5\pm0.5 \times 10^{10}M_{\odot}$ in what follows, with a corresponding estimate of $M_{10-40,{\rm M31}}$ three times smaller at $M_{10-40,{\rm M31}} \sim 5\pm2\times 10^{9}M_{\odot}$. 
The total stellar halo mass for the MW has been estimated to be between $M_{halo,{\rm MW}} = 4-7 \times 10^{8}M_{\odot}$ (\citealp{BHG16}, following \citealp{Bell08}); we assume a $M_{10-40,{\rm MW}} = 1.75\pm0.5 \times 10^{8}M_{\odot}$ three times smaller than the total stellar halo mass. There is evidence that the 3-D halo density profile changes slope from $-2.5$ to $-3.5$ at halo radii of around 25\,kpc \citep{Bell08,Sesar11,Xue15}, accordingly we assume a 3-D power law slope of $-3 \pm 0.5$. 
Following \citet{M16a}, the metallicity at 30 kpc of M31's halo is from \citet{Gilbert14}, correcting their values to an assumed alpha enhancement $[\alpha/{\rm Fe}] =0.3$.  The metallicity at 30 kpc of the MW's halo is the mean metallicity between the values reported in \citet{Sesar11} and \citet{Xue15}, i.e. ${\rm [Fe/H]} = -\textbf{}1.7$. Colour gradients for the MW and M31 are estimated from the metallicities at 10 and 40 kpc from \cite{Xue15} and \cite{Gilbert14}, respectively, using the \cite{Streich14} relationship between metallicity and RGB colour, assuming $[\alpha/{\rm Fe}] =0.3$. 

We note that resolved stellar populations data in the halo of the large S0 galaxy NGC 3115 \citep{Peacock15} exist along its minor axis, and at very large radii in the elliptical galaxy Centaurus A \citep{Rejkuba14}. 
We choose not to include these galaxies in our comparisons at this time owing to important differences between the halo profile and mass estimation techniques. In the case of NGC 3115, \cite{Peacock15} lack a measurement along NGC 3115's major axis, making it impossible to estimate the stellar halo axis ratio. Furthermore, Peacock et al.\ identify a low-metallicity tail with [Fe/H]$<-1$ as NGC 3115's stellar halo, whereas we identify all material at large minor axis radius as stellar halo. Our approach makes sense for galaxies dominated by a geometrically thin stellar disc, but may be less appropriate for a bulge-dominated galaxy such as NGC 3115. In the case of Centaurus A, \citet{Rejkuba14} derive no estimate of total stellar halo mass, and again it is unclear how to proceed with quantitative halo analysis in an elliptical galaxy, where one expects a large number of stars from the central parts of the galaxy to have been scattered to large radii during the violent relaxation that shapes the main body of the galaxy. We note that both galaxies have metallicities ${\rm [Fe/H]}_{30kpc}>-0.6$ and appear to have fairly substantial masses (in excess of $10^{10} M_{\odot}$) between 10 and 40kpc, so appear to be qualitatively consistent with the trends discussed here using the GHOSTS sample augmented with the Milky Way and M31. 

In panel $a$ of Fig.\ \ref{correlations} we show the ratio of the stellar halo mass to the total galaxy stellar mass, as a function of total galaxy stellar mass. This sample of Milky Way mass disc-dominated galaxies with total stellar masses between $4\times10^{10} M_{\odot}$ and $10^{11} M_{\odot}$ shows a remarkably large range of stellar halo mass fractions, varying by a factor of $\sim 7$ in GHOSTS galaxies alone, with variations of a factor of $\sim 15$ in stellar halo mass fractions with the addition of the Milky Way and M31 to the sample. 

In panel $d$ of Fig.\ \ref{correlations}, we show the inferred stellar halo  minor axis 3-D density power law slope as a function of inferred stellar halo mass between 10 and 40 kpc. The stellar haloes in nearby Milky Way mass disc galaxies have masses between 10 and 40 kpc that range between $\sim 10^{8.2}$ to $10^{9.7}$ M$\odot$, a {\it factor of 30} range in mass. These galaxies show a range in halo power law slopes between 10 and 40\,kpc, with 3D equivalent minor axis power law slopes between $-3$ and $-4.7$. Recall that many galaxies' density profiles have considerable deviations from power laws; we parameterise the haloes with power laws to facilitate comparison, cognizant that power laws are rarely accurate descriptions of the density profile of structured stellar haloes. 

We explore this issue more in Fig.\ \ref{surfacebrightness}, where we show the {\it projected} minor axis density profiles for the GHOSTS massive galaxy sample (datapoints without errorbars for clarity) together with the best-fit power laws. One can see that the stellar haloes of the GHOSTS Milky Way-mass galaxies are very broadly consistent with each other. While there is diversity in density at radii $<10$\,kpc, the scatter in halo densities appears to increase somewhat towards larger galactocentric radius. There may be a hint of a `minimum', relatively steep density profile (largely traced out by e.g., NGC 3031 or NGC 4565), with galaxies being able to have excursions to considerably higher density at a range of radii (e.g., the excess of NGC 4945 between 7 and 20kpc, tending towards lower values outside of 30kpc, or the dramatic density shelf in NGC 891, returning to values characteristic of NGC 3031 or NGC 4565 at $>40$\,kpc). This behavior may make intuitive sense --- one could easily imagine that at a given galaxy mass, a superposition of relatively low mass disrupted satellites may give a minimal `standard build' halo, whereas galaxies that managed to accrete one or more rather more massive satellites would augment this `standard build' halo, raising its density profile at a range of radii, and possibly enhancing substructure.

Panels $b$ and $e $  of Fig.\ \ref{correlations} show the metallicity of the stellar halo at 30\,kpc as a function of stellar halo mass between 10 and 40\,kpc (panel $b$) and rotation velocity (panel $e$). The stellar haloes of Milky Way mass disc galaxies have {\it an order of magnitude range in stellar halo median metallicity}, from $[\rm{Fe/H}]\sim -0.7$ dex to $-1.7$ dex. This halo metallicity appears not to correlate with rotation velocity (panel $e$), but does correlate strongly with {\it stellar halo} mass (panel $b$). The correlation between stellar halo metallicity and mass is the strongest correlation in our dataset, with a Pearson correlation coefficient of 0.89, corresponding to a 0.3\% chance of being drawn from an uncorrelated dataset.  
We will return to this correlation shortly.

Panels $c$ and $f$ of Fig.\ \ref{correlations}  show the minor axis stellar halo RGB colour gradient (a proxy for metallicity gradient) as a function of stellar halo mass between 10 and 40\,kpc (panel $c$) and rotation velocity (panel $f$). As highlighed in \citet{M16a}, Milky Way mass disc galaxies appear to have a range of behaviors in terms of their metallicity gradients, where $\sim$1/2 of the sample have no or a weak metallicity gradient, and the other half have strong metallicity gradients. The gradient does not appear to vary systematically with stellar halo mass. In our dataset, halo metallicity gradient appears tentatively to favour steeper negative gradients for galaxies with higher $v_{max}$, but with little statistical significance (Pearson correlation coefficient of $-0.62$, corresponding to a 10\% chance of being drawn from an uncorrelated dataset). 

\subsection{A Correlation Between Stellar Halo Metallicity and Stellar Halo Mass}
\label{sec:metmass}

\begin{figure}
	\includegraphics[width=75mm]{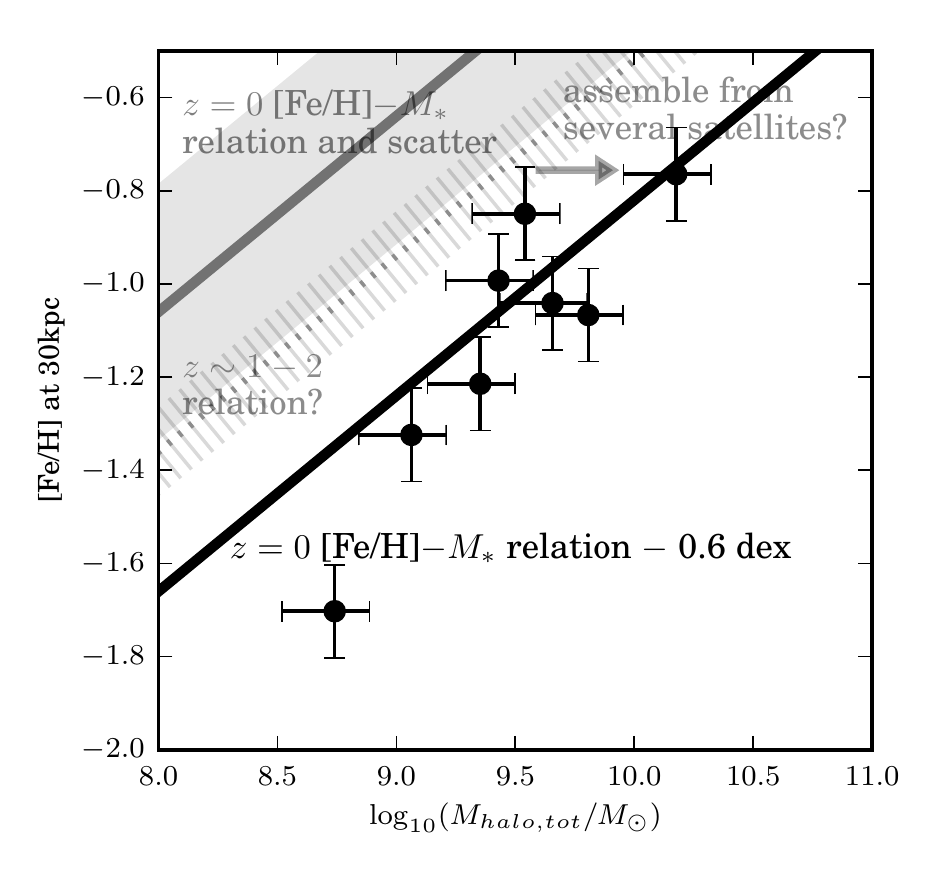}
	\caption{The [Fe/H]--$M_{halo,tot}$ relation for the GHOSTS galaxies, the Milky Way and M31. Total halo masses are estimated by multiplying the masses between 10 and 40\,kpc by a factor of 3. The [Fe/H] is characterised by the [Fe/H] at 30\,kpc along the minor axis. Overplotted is the $z=0$ galaxy stellar metallicity--stellar mass relation in grey (with $\sim 0.2$\,dex scatter), and a relation offset by $-0.6$ dex in black to give a rough estimate in the offset between the galaxy and stellar halo metallicity--mass relations. The dashed area is the $z\sim2$ gas metallicity--mass relation, shifted by $\sim -0.3$ dex from the present day relation. One can see that the relationship between stellar halo mass and metallicity has a shape that is broadly consistent with the galaxy metallicity--mass relation offset to slightly lower metallicities, as would be broadly expected if haloes are assembled from the debris of the disruption of multiple dwarf galaxies at intermediate redshifts (z=1-2). }
    \label{metmass}
\end{figure}

The most significant correlation between stellar halo properties in our dataset is a correlation between the mass of stellar haloes and their metallicity at 30\, kpc along the minor axis (panel $b$ of Fig.\ \ref{correlations}). In Fig.\ \ref{metmass}, we scale Fig.\ \ref{correlations} to `total' stellar halo masses, estimated by multiplying the $M_{10-40}$ values by a factor of 3, following \S \ref{BJ_Test}. A very rough fit to the data (using orthogonal distance regression) is: ${\rm [Fe/H]}_{30\,kpc} \sim (-1.45\pm0.1) + (0.7\pm0.15)(\log_{10} M_{halo,tot}-9)$; the slope is changed significantly by the Milky Way's stellar halo metallicity estimate, and is accordingly rather uncertain. In order to build intuition about how to interpret this relation, we show in Fig. \ref{metmass} the local galaxy stellar metallicity--stellar mass relation (obtained by tying together the stellar metallicity--stellar mass relations of \citealp{Kirby13} and \citealp{Gallazzi05}) in grey along with an assumed scatter of $\sim 0.2$\, dex which is comparable to the $<0.15$\,dex scatter of \citet{Kirby13} and the scatter at high stellar masses of \citet{Gallazzi05}. The black solid line is the local galaxy metallicity--mass relation offset by $-0.6$\,dex, which is scaled to go through the majority of the data points. The galaxy metallicity--mass relation is consistent with but slightly ($<2\sigma$) shallower than the stellar halo metallicity--mass relation.

We argue the broad similarity of the slope of the stellar halo metallicity--halo mass relation with the galaxy metallicity--mass relation is no coincidence, and that its offset is broadly as expected in a cosmological context. Accretion only models of halo formation in a cosmological context predict that most of the mass in stellar haloes should come from the few most massive progenitors { \citep[e.g.,][]{Deason16}}, and that the main epoch of halo building is at $z \sim 1-2$ \citep{Bullock01,BJ05,Cooper10}. As a (very rough) guide to what kind of offset one might expect at earlier times from the metallicity--mass relation,  \citet{Erb06} find that the $z\sim2$ {\it gas} metallicity--mass relation is offset by $\sim -0.3$ dex from the present day relation (shown as a dashed area in Fig.\ \ref{metmass}). 
Noting that most of the mass in stellar haloes should be formed from the disruption of the largest progenitors, the remaining offset should be interpreted as a largely {\it horizontal} offset (the arrow in Fig.\ \ref{metmass}), with lower mass galaxies being incorporated into a larger halo, where the relations are broadly consistent with the idea that the few largest progenitors provide most of the mass of a stellar halo at the present day, at a given metallicity characteristic of their metallicity at the time of accretion. While this argument is both necessarily rough and approximate, it is clear that there is an intuitive and quantitatively reasonable accretion-only framework in which to interpret the relation between the present-day stellar halo metallicity along the minor axis and the stellar halo mass {(see \citealp{Deason16} for an in-depth discussion of this issue)}. This accretion-only framework appears also consistent with recent results from hydrodynamical simulations, where the {\it in situ} halo component is expected to be negligible at  $R \ga 15$ kpc along the minor axis \citep{Pillepich15, M16b}. Thus, characterizing the halo metallicity with its metallicity along the minor axis yields a robust measurement of the properties  of the dwarf satellites that were tidally disrupted by the central galaxy.

\subsection{Comparison of Stellar Halo Mass Fractions with Observations} \label{sec:compobs}

\begin{figure}\centering
	\includegraphics[width=75mm]{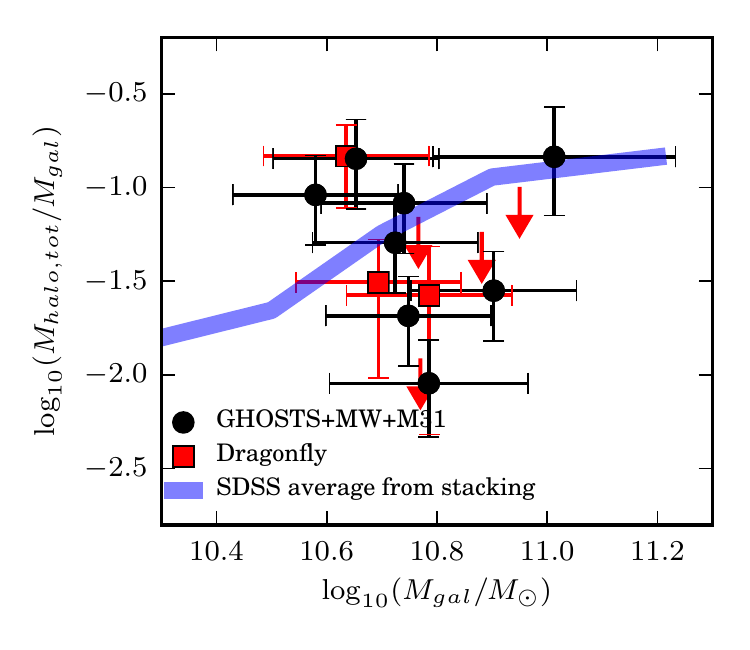}
	\caption{{A comparison of estimates of the fraction of stars in Milky Way mass galaxies that are in its stellar halo. Black points denote the measurements from GHOSTS from this paper, red datapoints show 3 estimates and 4 upper limits from \protect\citet{Merritt16}, and blue shows an estimate of average accreted fraction for low concentration galaxies using stacking of SDSS images \protect\citep[and priv.\ comm.]{Dsouza14}. }}
    \label{compobs}
\end{figure}

{A number of estimates or upper limits for stellar halo masses have recently become available from integrated light studies \citep{Dsouza14, Merritt16}. While there are no galaxies in common between the datasets, one can fruitfully compare the fraction of mass in stellar haloes for Milky Way mass galaxies between these studies.

In Fig.\ \ref{compobs}, we show the overall good level of agreement between estimated fractions of mass in stellar haloes from GHOSTS, MW, and M31 (this paper; black) and two other studies based on integrated light. 

Red symbols denote detections or upper limits of stellar haloes of galaxies with stellar masses above $4 \times 10^{10}M_{\sun}$ from deep imaging with the Dragonfly telescope array \citep{Merritt16}. We interpret galaxies with halo fraction error bars that intersect zero to be non-detections, and take the upper error bar from their Table 1 to be the upper limit to the halo mass fraction. An additional complication is that their estimated stellar halo mass fraction refers to light measured outside 5 half-mass radius $R_h$ only. We correct their estimates to an estimate of total stellar halo mass following \S \ref{BJ_Test}. We convert major axis $R_h$ into minor axis equivalent values using axis ratio estimates from the 2MASS Large Galaxy Atlas \citet{Jarrett03} when possible or from the NASA/IPAC Extragalactic Database. For most of their sample, $5R_h$ corresponds to a minor axis radius of $\sim 10$\,kpc, and we adopt the factor of 3 correction determined in \S \ref{BJ_Test} to scale their measurements to `total' mass. NGC 4220 has a smaller minor axis equivalent radius of $\sim 4$\,kpc, corresponding to a rough factor of two extrapolation to total mass. Merritt et al.'s imaging for M101, owing to its low inclination and large disk extent, is sensitive to a stellar halo only outside of $\sim 30$\,kpc, and following \S \ref{BJ_Test} we estimate a factor of ten extrapolation to total stellar mass. The adopted stellar halo mass fractions or limits are tabulated in Table \ref{dragonfly}. We find that the treatment of Dragonfly non-detections as upper limits, and a model-motivated extrapolation to total stellar halo fraction brings the measurements of \citet{Merritt16} into line with ours. These studies paint a consistent picture of there being around a factor of $\sim 15$ full range in stellar halo mass fractions for roughly Milky Way mass galaxies, with the Dragonfly non-detections being clustered towards the lower side of our observed range of stellar halo fractions.

Estimates of the average accreted stellar mass fraction for low-concentration disk-dominated galaxies from stacking of the SDSS
\citep{Dsouza14} 
are given in blue. 
\citet{Dsouza14} give an estimate of the outer light fraction; we adjust it downwards by 0.1 dex to account for stellar M/L differences between the bluer outer envelopes of galaxies and their redder, higher stellar M/L cores, using a typical color difference of $\Delta (g-r) \sim 0.1$ from 
\citet{Dsouza14} and estimates of stellar M/L from \citet{Bell03}. Furthermore, D'Souza (private communication) has found using stacking of mock images from the Illustris hydrodynamical simulations \citep{Vogelsberger14} that the outer light fraction is $\sim 0.1$\,dex systematically higher than the fraction of accreted stars in those simulations. Accordingly, following D'Souza's recommendation, we adjust their accreted light fraction down by another 0.1 dex. This line, denoted in blue, represents the best available estimate from SDSS stacking of the accreted fraction of stars in low-concentration galaxies. This average is in good accord with the typical stellar halo fractions measured by GHOSTS or Dragonfly. By focusing on individual halos, GHOSTS and Dragonfly enrich the results of \citet{Dsouza14} by showing that galaxies have considerable scatter in stellar halo mass around that relation, with a full order of magnitude or more range in halo mass fraction at stellar masses comparable to the Milky Way.

}

\subsection{Comparison Between Stellar Halo Observations and Models of Stellar Halo Formation} \label{sec:comp_models}

\begin{figure*}\centering	
	\includegraphics[width=170mm]{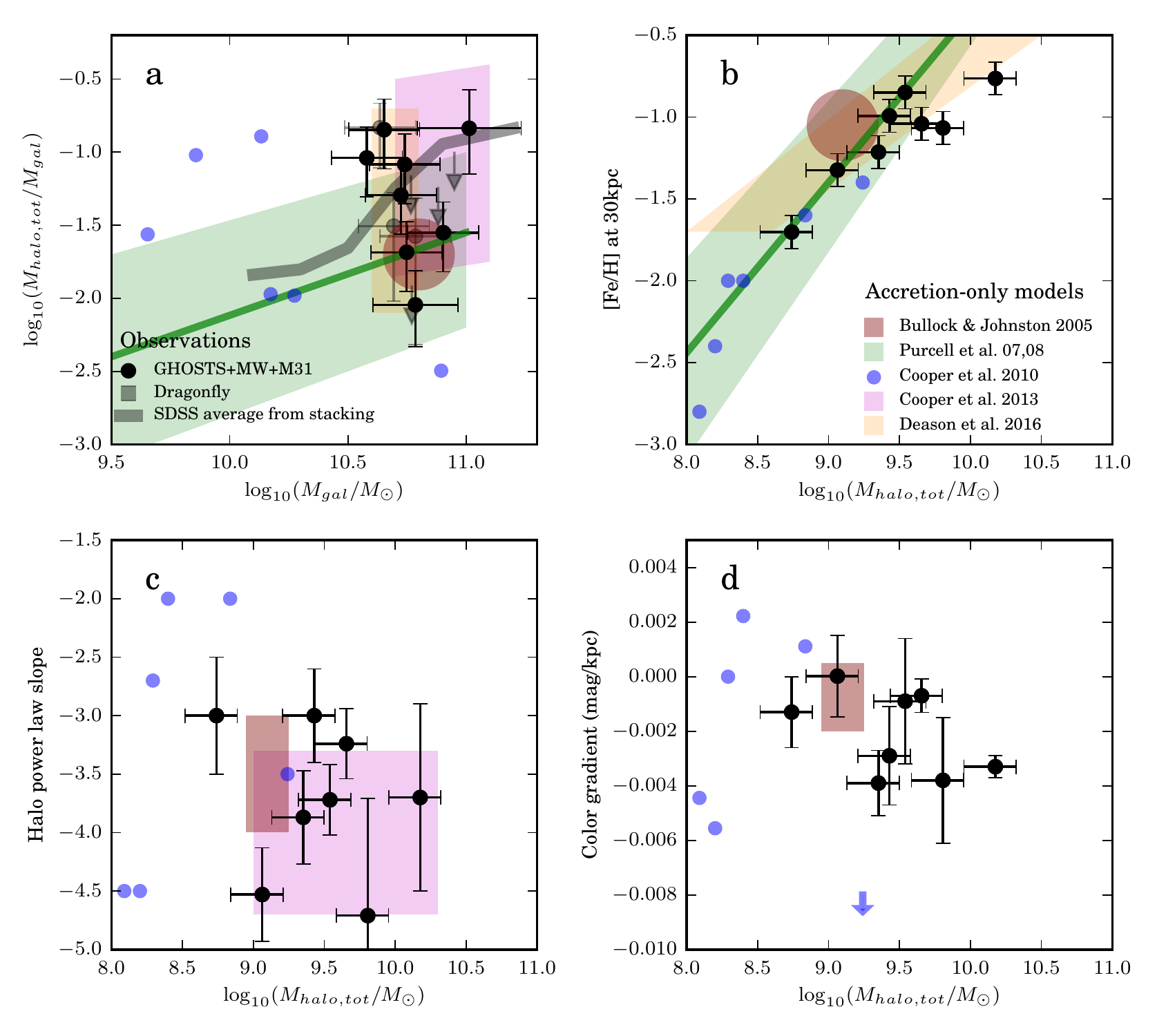}
	\caption{Comparison to accretion-only models --- Panel $a$: ratio of `total' stellar halo mass and total stellar mass, as a function of total stellar mass. Panel $b$: stellar halo metallicity at 30kpc as a function of `total' stellar halo mass. Panel $c$: inferred 3-D stellar halo density power law slope as a function of `total' stellar halo mass. Panel $d$: stellar halo colour gradient (a proxy for metallicity gradient) as a function of `total' stellar halo mass. The observational data are shown in black and grey. Models: {\it brick red area}: \protect\cite{BJ05}, {\it light green$+$line}: \citet{Purcell07,Purcell08}, {\it blue}: \protect\cite{Cooper10}, {\it magenta}:  \citet{Cooper13}, {\it orange}: \citet{Deason16}.}
	\label{model_acc}
\end{figure*}

\begin{figure*}\centering	
	\includegraphics[width=170mm]{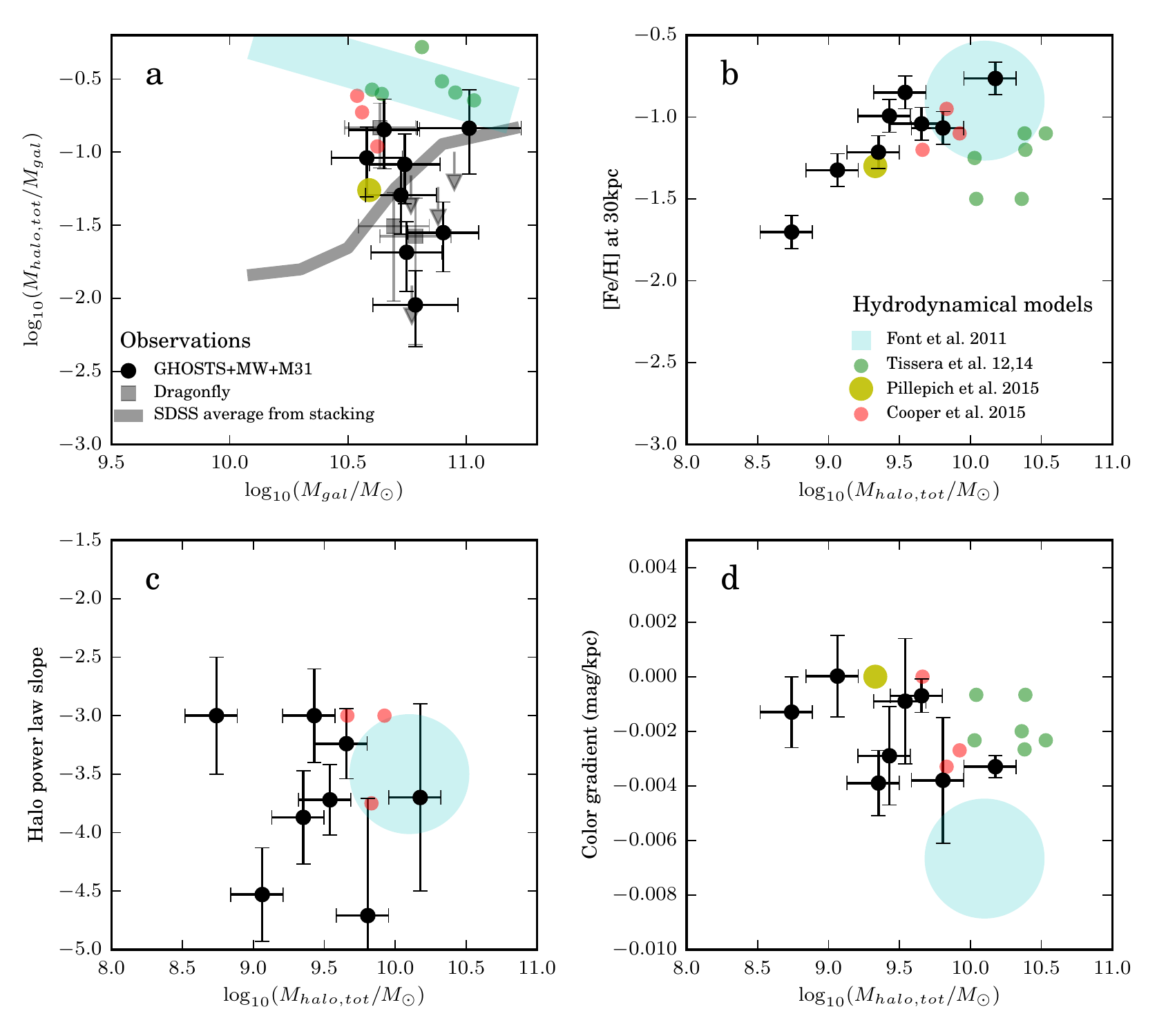}
	\caption{Comparison to hydrodynamical models --- Panel $a$: ratio of `total' stellar halo mass and total stellar mass, as a function of total stellar mass. Panel $b$: stellar halo metallicity at 30kpc as a function of `total' stellar halo mass. Panel $c$: inferred 3-D stellar halo density power law slope as a function of `total' stellar halo mass. Panel $d$: stellar halo colour gradient (a proxy for metallicity gradient) as a function of `total' stellar halo mass. The observational data are shown in black and grey. Hydrodynamical models: {\it light cyan}: \protect\citet{Font11}, {\it green}: \citet{Tissera12,Tissera14}, {\it yellow}: \protect\cite{Pillepich15}, {\it red}: \citet{Cooper15}. }
	\label{model_hydro}
\end{figure*}

Given the diversity of halo properties seen in Figs.\ \ref{correlations} and \ref{surfacebrightness}, and the strength of the correlation between halo metallicity along the minor axis and stellar halo mass seen in Figs.\ \ref{correlations} and \ref{metmass}, we turn now to a preliminary and relatively rudimentary comparison between these observational results with expectations from models of galaxy formation in a cosmological context. We make this comparison as fair as possible, noting that the majority of the predicted quantities, such as metallicity and metallicity gradients, are typically presented as spherically averaged properties in models whereas we measure them as projected along the disc minor axis, thus some discrepancies may arise due to the different methodology used. We note that models of stellar halo formation in a cosmological context is a rapidly developing field; in this spirit, we chart out some broad patterns and ideas, but leave detailed comparisons to future works and works by those modeling stellar halo formation in a cosmological context. In what follows, we focus on comparing expectations and predictions for the stellar haloes of disc galaxies with roughly the Milky Way's total (dark plus baryonic) mass of  $0.5 - 2\times 10^{12}M_{\odot}$ with our observations, cognizant of the considerable difficulty in measuring the dark halo mass of our and nearby galaxies.

\subsubsection{Accretion-only models}

We first compare with models which account for the build up of stellar haloes from the tidal disruption of dwarf satellites only (accretion-only models) in Fig.\ \ref{model_acc}. Light green shaded areas denote the region occupied by roughly 95\% of the accretion-only haloes from the halo occupation models of \citet{Purcell07, Purcell08}. The \citet{Purcell08} metallicity--stellar halo mass relation is approximated using their Fig.\ 7, assuming that their galaxies in $\sim 10^{12} M_{\odot}$ haloes have stellar masses $\sim 10^{10.75} M_{\odot}$. In order to convert metallicity into [Fe/H], we assume [Fe/H]$_{Purcell} \sim \log_{10} Z/Z_{\odot} - 0.2$, assuming $\rm [\alpha/Fe] \sim 0.3$. Brick red denotes the results from the \citet{BJ05} models (see also \citealp{Font06a}); these haloes have a narrow $\sim \times 2$ range in stellar mass. Blue symbols denote individual haloes from high resolution $N$-body simulations by \citet{Cooper10}, where we again estimate iron abundance by subtracting 0.2\,dex from the predicted metallicity. {Magenta regions enclose the 68\% range of the structural properties of haloes expected to host galaxies with disks ($B/T<0.9$) modeled using the Millennium-II $N$-body simulations by \citet{Cooper13}. Orange regions enclose the range of predictions from \citet{Deason16}.} 

Accretion-only models, for the most part, predict a reasonably realistic range of stellar halo masses (panel $a$ of Fig.\ \ref{model_acc}). Similarly, the broad range of power-law slopes of the accretion-only models seen in panel $c$ of Fig.\ \ref{model_acc} appears to be in reasonable accord with the data, with the possible exception of the \citet{Cooper10} models, which may have too wide a diversity in halo density profiles. Panel $d$ of Fig.\ \ref{model_acc} suggests that the weakness and uniformity of metallicity gradients of the \citet{BJ05} models is in conflict with the observations --- the variability of stellar halo metallicity gradients from galaxy to galaxy in the \citet{Cooper10} models appears to match more accurately the observational data. 

Most notably, and as foreshadowed in \S \ref{sec:metmass}, the accretion-only models predict a strong stellar halo mass--metallicity relation, in striking accord with the data. While the form of the relation was argued to be inevitable, given the stellar mass--stellar metallicity relation of the progenitor satellites (which is typically baked into the accretion-only models as a constraint) and given the almost negligible contribution from {\it in situ} halo stars at the radius along the minor axis where the [Fe/H] was derived (see \citealt{Pillepich15, M16b}), {the normalisation of the relation compared to the galaxy metallicity--mass relation is less trivial and appears to have been predicted accurately by the models. This match in normalization suggests that the accretion-only models have assembled their stellar haloes from appropriate progenitors with reasonable metallicities at the time of satellite infall.} Note that a broad relation between halo metallicity and total {\it galaxy} luminosity was suggested by \cite{Mouhcine05b}; the observations (and models) predict that a broad relation of this type is expected, but it is driven by a much more fundamental stellar halo mass--metallicity relation --- bigger stellar haloes are assembled from bigger, more metal rich pieces \citep{Deason16}. 

{This overall impressive level of agreement is encouraging. On one hand, the disagreements are now subtle enough that differences between observational and model metrics will matter, motivating an effort by modelers and observers to compare more consistently with each other in an effort to refine the models and better interpret the observations. On the other hand, this agreement motivates the use of stellar haloes to {\it quantitatively} explore galactic merger histories: \citet{Deason16} show that there is a relationship between the stellar halo mass, metallicity and the properties of the most massive progenitor, suggesting that we can gain precious insight into the most massive mergers that affected many of our nearby galactic neighbours.  }

\subsubsection{Hydrodynamical models}

We now compare with hydrodynamical models, which account for both the accreted plus {\it in situ} build up of stellar haloes, in Fig.\ \ref{model_hydro}.
Hydrodynamical models attempt to more completely capture the gas, stellar and feedback physics of galaxy formation; they are both complex and computationally expensive.  Importantly, hydrodynamical simulations also predict that an {\it in situ} stellar halo should exist, with contributions from both early chaotically-distributed star formation during the assembly of the galaxy and stars kicked up from deeper in the potential well (see \citealt{Cooper15} for a thoughtful discussion of the origin of the {\it in situ} stellar halo). These {\it in situ} haloes vary from model to model in prominence (\citealt{Cooper15} emphasise the importance of different choices for sub-grid physics on the properties of the {\it in-situ} halo), but typically carry $>5 \times 10^9 M_{\odot}$ \citep{Zolotov09,Pillepich15, Cooper15}. The {\it in-situ} halo appears to show a metallicity difference from the accreted stars in most models, although both components have broad distributions and the distribution of {\it in situ} metallicities is likely to be sensitive to input physics (see, e.g., Fig.\ 6 in \citealt{Cooper15} and the illuminating and forthright discussion in \citealt{Zolotov10}). \citet{Pillepich15} and \citet{M16b} emphasise that the {\it in situ} halo may be highly flattened, and their metallicity signatures may be difficult to discern along the minor axis. 

The broad properties of hydrodynamical models vary considerably from model to model (Fig.\ \ref{model_hydro}), but the models appear for the most part to overpredict stellar halo masses (panel $a$; e.g., light cyan for \citealp{Font11}, green for \citealp{Tissera12, Tissera14}, red for \citealp{Cooper15}, and yellow for \citealp{Pillepich15}). 
This overprediction of stellar halo mass seems particularly acute --- nearly an order of magnitude --- for lower resolution hydrodynamical simulations. While we show examples from \citet{Font11} and \citet{Tissera12, Tissera14}, similar behavior is seen in the simulated haloes of e.g., \citet{Zolotov09}, \citet{Bailin14} and \citet{Pillepich14}. Such a large discrepancy implies that both the accreted and {\it in situ} parts of the halo are overproduced in such lower-resolution models, as both components separately violate observational constraints. 
Substantially higher resolution more current simulations \citep{Cooper15,Pillepich15} have stellar haloes that are in rather better accord with the observations, and appear to overpredict halo mass by a factor of less than three, where systematic differences in how stellar haloes are defined between simulations and the observations may play an important role (see, e.g., the discussion in \citealp{Pillepich15}). 

Hydrodynamical model stellar halo metallicites at 30\,kpc (panel $b$ of Fig.\ \ref{model_hydro}) are similar to the observational estimates at [Fe/H]$\sim -1.2$, but would be rather low for their halo stellar mass. Lower-resolution hydrodynamical models (cyan and green) do not reproduce the observed strong correlation between stellar halo mass and stellar halo metallicity. More current, higher-resolution simulations (red and yellow) have similar metallicities, but owing to their lower and somewhat more realistic stellar masses lie closer to the observed galaxies in the stellar halo mass--stellar halo metallicity plane. With only 4 simulations to explore, it is too early to say if high-resolution hydrodynamical simulations naturally reproduce the observed stellar halo metallicities--stellar halo mass correlation. Like \citet{Pillepich15} and \citet{M16b}, we caution that with the moderate level of discrepancy now seen between high-resolution hydrodynamical models and the observations the choice of observational metric matters; accordingly, it will be important to estimate halo masses and metallicities of the simulated haloes in ways that connect well with these observations.

Only two simulations predict halo power law slopes, and these simulations broadly match the observational constraints (panel $c$). In addition, and not shown, their axis ratios tend to be $c/a \sim 0.6$ but with considerable range in axis ratio (e.g., \citealt{McCarthy12} for the \citealp{Font11} models), in reasonable agreement with the observations. 

Finally, a diversity of stellar halo metallicity gradients were predicted by current high-resolution hydrodynamical simulations --- \citet{Cooper15} and \citet{Pillepich15} --- and appear to be in accord with the data (panel $d$ of Fig\ \ref{model_hydro}). The simulations of \citet{Tissera12,Tissera14} also reproduce the observed diversity of metallicity gradients, remembering that their overall stellar halo masses are generally dramatically overpredicted. Observations appear to rule out ubiquitous and large metallicity gradients of the kind predicted by \citet{Font11}. We caution that our minor-axis metallicity gradients may be relatively insensitive to exploring the fraction of stars from {\it in situ} formation of halo stars; instead, major axis metallicity gradients may prove a more decisive probe of the importance of {\it it situ} stars in stellar haloes, owing to their predicted high degree flattening in current high-resolution simulations \citep{Pillepich15,M16b}.

\subsubsection{Degree of substructure}

While metrics of the number and prominence of individual streams or shells have been proposed and discussed \citep[e.g.,][]{Johnston08,Atkinson13,Hendel15,Amorisco15a}, fewer works have quantified the degree of substructure of the `aggregate' stellar halo \citep[e.g.,][]{Bell08,Amorisco15}.
Our prime observational measure of the degree of substructure is the intrinsic scatter that must be included to our power law model in order to make it an acceptable fit to the data, similar in spirit to the RMS/Total estimates of \citet{Bell08} and the tidal parameter estimates of elliptical galaxies by \citet{Tal09}.  For our sample, the typical intrinsic scatter around the fit is $0.05-0.1$dex, ranging from undetectable to 0.14 dex for the highly structured major axis profile of NGC 3031. 

This metric for the degree of substructure of the `aggregate' halo has been very rarely calculated by modelers. Our own fits to the \citet{BJ05} simulations (hybrid disc$+$bulge$+$dark halo potential plus $N$-body satellites) yield values of $\sim 0.1$\,dex, in broad accord with the observations. We do note that it is likely that some models would fail to match the observational constraints: the models of \citet{Cooper10} have strong substructure, far in excess of that seen in the Milky Way or in the \citet{BJ05} models (compare the RMS/total measures of \citealt{Helmi11}  with \citealp{Bell08}). \citet{Bailin14} attribute the high degree of substructure in pure $N$-body only models (which display triaxial, very structured stellar haloes) to the lack of a potential from the main body of the galaxy; the potential from the main body of the galaxy appears to lead to precession which erases substructure and produces a more oblate halo, in better accord with the data. We strongly encourage simulators to produce quantitative estimates of `aggregate' stellar halo subtructure, as a crucial test of their input model physics.

\section{Summary and Conclusions}
\label{sec6}

We have examined the halo stellar populations of six galaxies from the GHOSTS survey \citep{RadSmith11}. HST/ACS  and HST/WFC3 data were used from fields observed along the major and minor axes of each halo. We construct CMDs from these observations and  select RGB stars above the 50\% completeness limit to trace the stellar halo populations in these galaxies. We use the selected RGB stars to derive a stellar density profile for each halo. From the density profiles we estimate a best-fit power-law slope and intrinsic scatter around a smooth power law, axis ratio, and stellar mass of each halo. 

We find a diversity of stellar halo masses between minor axis equivalent $10-40$\,kpc of $3-21 \times 10^{8} M_{\odot}$ and projected power law slopes of between $-2$ and $-3.7$ along the minor axes. Owing to substructure in stellar haloes (particularly prominent for example along NGC 891's minor axis), we measure a typical intrinsic scatter around a smooth power law fit of $0.05-0.1$ dex. By comparing the densities of the minor axis and major axis profiles for each galaxy at distances around $\sim 25$\,kpc, we infer axis ratios at $\sim 25$\,kpc ranging from $0.4-0.75$. 

Using the 11 halo realisations from the \citet{BJ05} models as a guide for interpreting a richly structured halo using sparse pointings, we estimate systematic uncertainties for the inferred stellar halo masses, power law slopes, and projected axis ratios for the GHOSTS galaxies. In particular, using the stellar halo mass measurements within $10-40$\,kpc for the \citet{BJ05} models and comparing them to the total stellar halo mass, we expect that the above $10-40$\,kpc halo masses should be around $30-40$\% of the total stellar halo mass for an accretion-dominated halo. Consequently, we expect that the GHOSTS galaxies have `total' stellar masses of around $1-6 \times 10^{9} M_{\odot}$.

In conjunction with measurements for the stellar halo properties of the Milky Way and M31, we find that the GHOSTS stellar haloes lie in between the extremes charted out by the (rather atypical) haloes of the Milky Way and M31. Galaxies with stellar masses similar to the Milky Way have an order of magnitude range in stellar halo mass, factors of several differences in characteristic minor axis halo metallicities, power-law profiles with best fit slopes varying between $-2$ and $-3.7$, and a variety of metallicity gradients, where $\sim 1/2$ of the sample have little to no measurable metallicity gradient. The sample shows a strong correlation between stellar halo metallicity and stellar halo mass. 

We compare our observational results with the results of models of stellar halo formation in a cosmological context. We find good agreement between accretion-only models, where the stellar haloes are formed by the disruption of dwarf satellites, and the observations. In particular, the strong observed correlation between stellar halo metallicity and stellar halo mass is naturally reproduced by the models as the result of a strong metallicity--mass relation of the satellite progenitors, plus the tendency for more massive stellar haloes to have been formed by the disruption of larger progenitors.
Low-resolution hydrodynamical models have unrealistically high stellar halo masses. Current high-resolution hydrodynamical models predict stellar halo masses somewhat higher than observed but in better accord with the data, with reasonable metallicities, metallicity gradients, and density profiles. The level of the differences between predictions and observations may be small enough that differences in definition between our observational and model metrics may be important.

\section*{Acknowledgments}
{We thank the referee for their helpful comments and suggestions}. We appreciate helpful conversations and feedback and insights from Sarah Loebman, Monica Valluri, Andrei Kravtsov, Nicolas Martin, and Oleg Gnedin. 
This work was supported
by NSF grant AST 1008342 and HST grants GO-11613 and GO-12213 provided by NASA through a grant from the Space Telescope Science Institute, which is operated by the Association of Universities for Research in Astronomy, Inc., under NASA contract NAS5- 26555. Additionally, some of the data presented in this paper were obtained from the Mikulski Archive for Space Telescopes (MAST). STScI is operated by the Association of Universities for Research in Astronomy, Inc., under NASA contract NAS5-26555. Support for MAST for non-HST data is provided by the NASA Office of Space Science via grant NNX09AF08G and by other grants and contracts.
We acknowledge the usage of the HyperLeda database (\url{http://leda.univ-lyon1.fr}).
This research has made use of the NASA/IPAC Extragalactic Database (NED) which is operated by the Jet Propulsion Laboratory, California Institute of Technology, under contract with the National Aeronautics and Space Administration.
This
work has made use of the {\sc iac-star} Synthetic CMD computation
code. {\sc iac-star} is supported and maintained by the IAC’s IT Division.
This work used the astronomy \& astrophysics package for
MATLAB \citep{Ofek14}. This research has made use of NASA's Astrophysics Data System Bibliographic Services. This research made use of Astropy, a community-developed core Python package for Astronomy \citep{astropy}.

\bibliographystyle{mnras}
\bibliography{stellhalos}

\begin{landscape}
\begin{table}
	\renewcommand{\arraystretch}{1.5}
	\begin{center}
		\caption{Table of Values for GHOSTS Galaxies}  
		\begin{tabular}{lrrrrrrrr}	
			\hline\hline
			&&NGC 253 &NGC 891 &NGC 3031 &NGC 4565 &NGC 4945 &NGC 7814 \\
			&Assumed Distance$^{**}$ (Mpc)&3.5 & 9.2 & 3.6 & 11.4 & 4.0 & 14.4\\
			\hline
			\multirow{3}{*}{Minor Axis}
			&$r_{0}$  (kpc)& 19.2& 23.3& 23.1& 22.9& 16.0&20.0\\
			&log$_{10}\Sigma_0$ (log$_{10}\ N\cdot$ arcsec$^{-2})$&$-1.73_{-0.02}^{+0.02}$  &$-2.06_{-0.04}^{+0.04}$  &$-2.63_{-0.02}^{+0.02}$        &$-2.11_{-0.02}^{+0.02}$        &$-1.94_{-0.03}^{+0.03}$        &$-1.39_{-0.07}^{+0.04}$&\\
			&Power law slope $\alpha\ (\pm 0.2)$&$-2.24_{-0.06}^{+0.07}$ &$-2.00_{-0.23}^{+0.33}$ &$-3.53_{-0.15}^{+0.18}$       &$-2.87_{-0.07}^{+0.08}$       &$-2.72_{-0.17}^{+0.16}$       &$-3.71_{-0.09}^{+0.99}$&\\
			&Intrinsic scatter $\sigma\ (\pm 0.03)$&$0.10_{-0.01}^{+0.01}$&$0.13_{-0.05}^{+0.05}$&$0.03_{-0.02}^{+0.02}$ &$< 0.11$ (95\%)&$0.05_{-0.02}^{+0.01}        $&$<0.03^{***}$&\\
			\hline
		    \multirow{3}{*}{Major Axis$^*$}
		    &$r_{0}$  (kpc)& 25.5& 26.7& 31.2& 40.7& 30.4&35.0\\
			&log$_{10}\Sigma_0$ (log$_{10}\  N\cdot$ arcsec$^{-2})$&$-1.34_{-0.02}^{+0.02}$   &$-1.91_{-0.03}^{+0.03}$        &$-2.34_{-0.06}^{+0.08}$        &$-1.92_{-0.05}^{+0.04}$        &$-1.94_{-0.02}^{+0.03}$        &$-1.53_{-0.13}^{+0.10}$&\\
			&Power law slope $\alpha\ (\pm 0.2)$&$-3.01_{-1.55}^{+1.09}$  &$-2.77_{-0.44}^{+0.73}$       &$-3.11_{-0.48}^{+0.88}$       &$-5.28_{-0.45}^{+0.47}$       &$-2.73_{-0.23}^{+0.23}$       &$-5.33_{-0.57}^{+3.34}$&\\
			&Intrinsic scatter $\sigma\ (\pm 0.03)$&$<0.05$ (95\%)  &$<0.03^{***}                        $&$0.14_{-0.06}^{+0.04}$        &$<0.03^{***}                         $&$0.09_{-0.02}^{+0.01}        $&$<0.03^{***}$&\\
			
			\hline
			&Projected $c/a_{25kpc}$ Axis Ratio $(\pm 0.1)$&$0.55_{-0.05}^{+0.04}$&$0.74_{-0.05}^{+0.04}$&$0.61_{-0.05,-0.2}^{+0.03,+0.0}$&$0.42_{-0.01}^{+0.02}$&$0.52_{-0.02}^{+0.02}$&$0.59_{-0.05}^{+0.14}$&\\

			&Stellar Halo Mass ($M_{10-40}$)($\pm 30$\%)&$1.45_{-0.10}^{+0.17}\times10^{9}$&$8.58_{-0.50}^{+0.72}\times10^{8}$&$3.66_{-0.22}^{+0.35}\times10^{8}$&$7.16_{-0.31}^{+0.33}\times10^{8}$&$1.11_{-0.06}^{+0.07}\times 10^{9}$&$2.05_{-0.26}^{+0.43}\times 10^{9}$&\\
            
            &Total Stellar Halo Mass ($M_{halo}$)($\pm 43$\%)&$4.53_{-0.31}^{+0.53}\times10^{9}$&$2.69_{-0.16}^{+0.23}\times10^{9}$&$1.14_{-0.07}^{+0.10}\times10^{9}$&$2.24_{-0.10}^{+0.10}\times10^{9}$&$3.47_{-0.22}^{+0.19}\times10^{9}$&$6.41_{-0.81}^{+1.34}\times10^{9}$&\\

			&Total Galaxy Stellar Mass ($M_{galaxy}$)&$5.5\pm{1.4}\times10^{10}$&$5.3\pm{1.3}\times 10^{10}$&$5.6\pm{1.4}\times 10^{10}$&$8.0\pm{2.0}\times 10^{10}$&$3.8\pm{0.95}\times 10^{10}$&$4.5\pm{1.1}\times 10^{10}$&\\

			&$V_{rot}$ (km$\cdot$ sec$^{-1}$)&194&212&224&245&167&231\\
			
			&Colour gradient$^{**}$ ($\times10^{-4}$ mag/kpc)&$-7.0\pm6.1$&$-29\pm18$&$0.2\pm15$&$-39\pm12$&$-9.0\pm23$&$-38\pm23$\\
			\hline
			
		\end{tabular} \label{Table_Slopes_Masses} \\
		$^*$Not used explicitly to estimate stellar halo mass.\\
				$^{**}$From \cite{M16a}. \\
				$^{***}$Three $\sigma$ upper limit based on the formal fit uncertainties. \\
                Systematic uncertainties are given in parentheses in the second column for the power law slope $\alpha$, the intrinsic scatter estimate $\sigma$, the axis ratio and estimates of stellar halo mass.\\
                The axis ratio for NGC 3031 has an additional asymmetric error as a result of the galaxy not being completely edge-on.\\
	\end{center}
\end{table}
\end{landscape}

\definecolor{grey}{rgb}{0.93,0.93,0.93}
\begin{table}
\tiny
	\begin{center}
		\caption{Star count Data for the Six GHOSTS Galaxies Examined in this Paper. Shading is included to differentiate between GHOSTS fields \label{tab:datapoints}}  
		\begin{tabular}{p{1.1cm} p{0.6cm} r r p{1.1cm} p{0.9cm} p{0.9cm}}
			\hline\hline
			Galaxy          & Axis  & Field  & r (kpc) & N$\cdot$Arcsec$^{-2}$ & $-1 \sigma$ & $+1 \sigma$ \\ \hline
			NGC 253         & Major & 7      & 23.77   & 0.0456                & 0.0415      & 0.0496      \\
			                &       &        & 24.45   & 0.0542                & 0.0515      & 0.0570      \\
			                &       &        & 25.17   & 0.0546                & 0.0522      & 0.0569      \\
			                &       &        & 25.92   & 0.0431                & 0.0409      & 0.0452      \\
			                &       &        & 26.63   & 0.0378                & 0.0356      & 0.0401      \\
			                &       &        & 27.37   & 0.0343                & 0.0309      & 0.0377      \\
			\rowcolor{grey} & Minor & 8      & 7.11    & 0.131                 & 0.127       & 0.136       \\
			\rowcolor{grey} &       &        & 7.69    & 0.114                 & 0.110       & 0.118       \\
			\rowcolor{grey} &       &        & 8.28    & 0.0955                & 0.0919      & 0.0992      \\
			\rowcolor{grey} &       &        & 8.89    & 0.0689                & 0.0657      & 0.0720      \\
			\rowcolor{grey} &       &        & 9.51    & 0.0693                & 0.0661      & 0.0724      \\
			\rowcolor{grey} &       &        & 10.03   & 0.0549                & 0.0516      & 0.0581      \\
			                &       & 9      & 4.44    & 0.567                 & 0.556       & 0.578       \\
			                &       &        & 5.03    & 0.450                 & 0.443       & 0.457       \\
			                &       &        & 5.67    & 0.309                 & 0.303       & 0.315       \\
			                &       &        & 6.38    & 0.246                 & 0.240       & 0.251       \\
			                &       &        & 7.03    & 0.224                 & 0.219       & 0.229       \\
			                &       &        & 7.59    & 0.201                 & 0.194       & 0.209       \\
			\rowcolor{grey} &       & (W) 11 & 17.56   & 0.0225                & 0.0192      & 0.0259      \\
			\rowcolor{grey} &       &        & 18.08   & 0.0229                & 0.0207      & 0.0250      \\
			\rowcolor{grey} &       &        & 18.57   & 0.0210                & 0.0191      & 0.0229      \\
			\rowcolor{grey} &       &        & 19.16   & 0.0196                & 0.0178      & 0.0214      \\
			\rowcolor{grey} &       &        & 19.72   & 0.0209                & 0.0190      & 0.0229      \\
			\rowcolor{grey} &       &        & 20.18   & 0.0178                & 0.0151      & 0.0206      \\
			                &       & 12     & 23.03   & 0.0096                & 0.00787     & 0.0114      \\
			                &       &        & 23.73   & 0.0105                & 0.00933     & 0.0118      \\
			                &       &        & 24.31   & 0.0097                & 0.00860     & 0.0107      \\
			                &       &        & 25.00   & 0.0103                & 0.00922     & 0.0114      \\
			                &       &        & 25.67   & 0.0100                & 0.00885     & 0.0111      \\
			                &       &        & 26.31   & 0.0115                & 0.00970     & 0.0134      \\
			\rowcolor{grey} &       & 13     & 31.29   & 0.00879               & 0.00756     & 0.0100      \\
			\rowcolor{grey} &       &        & 31.83   & 0.0104                & 0.00915     & 0.0116      \\
			\rowcolor{grey} &       &        & 32.39   & 0.00753               & 0.00647     & 0.00860     \\
			\rowcolor{grey} &       &        & 33.00   & 0.00623               & 0.00525     & 0.00721     \\
			\rowcolor{grey} &       &        & 33.59   & 0.00809               & 0.00698     & 0.00920     \\
			\rowcolor{grey} &       &        & 34.11   & 0.00721               & 0.00610     & 0.00832     \\
			                &       & (W) 14 & 37.70   & 0.00347               & 0.00304     & 0.00370     \\
			\rowcolor{grey} &       & (W) 15 & 44.71   & 0.000872              & 0.000579    & 0.00117     \\
			                &       & 16     & 49.49   & 0.000932              & 0.000662    & 0.00120     \\
			\rowcolor{grey} &       & (W) 17 & 53.44   & 0.000731              & 0.000444    & 0.00102     \\
			                &       & 18     & 58.33   & 0.000718              & 0.000457    & 0.000978    \\
			\rowcolor{grey} &       & (W) 19 & 70.90   & 0.00103               & 0.000721    & 0.00134     \\
			                &       & 20     & 75.72   & 0.000852              & 0.000586    & 0.00112     \\ \hline
			NGC 891         & Major & 12     & 21.13   & 0.0224                & 0.0213      & 0.0235      \\
			                &       &        & 25.78   & 0.0132                & 0.0124      & 0.0140      \\
			\rowcolor{grey} &       & 13     & 28.44   & 0.0118                & 0.0110      & 0.0126      \\
			\rowcolor{grey} &       &        & 32.67   & 0.00594               & 0.00540     & 0.00648     \\
			                & Minor & 1      & 6.96    & 0.138                 & 0.135       & 0.141       \\
			                &       &        & 9.27    & 0.0515                & 0.0488      & 0.0541      \\
			\rowcolor{grey} &       & 5      & 24.27   & 0.00722               & 0.00640     & 0.00804     \\
			\rowcolor{grey} &       &        & 26.62   & 0.00695               & 0.00617     & 0.00773     \\
			\rowcolor{grey} &       &        & 29.35   & 0.00701               & 0.00623     & 0.00779     \\
			\rowcolor{grey} &       &        & 31.44   & 0.00635               & 0.00549     & 0.00720     \\
			                &       & 6      & 14.53   & 0.0164                & 0.0145      & 0.0182      \\
			                &       &        & 17.29   & 0.0141                & 0.0131      & 0.0152      \\
			                &       &        & 20.12   & 0.0102                & 0.00942     & 0.0109      \\
			                &       &        & 22.87   & 0.00598               & 0.00495     & 0.00700     \\
			\rowcolor{grey} &       & (W) 7  & 33.26   & 0.00438               & 0.00381     & 0.00495     \\
			\rowcolor{grey} &       &        & 37.15   & 0.00705               & 0.00633     & 0.00776     \\
			                &       & 8      & 47.39   & 0.000376              & 0.000206    & 0.000546    \\
			                &       &        & 52.60   & 0.000106              & 0.000010    & 0.000214    \\ \hline
			NGC 3031        & Major & 4      & 22.12   & 0.0106                & 0.00944     & 0.0117      \\
			                &       &        & 23.49   & 0.00852               & 0.00785     & 0.00919     \\
			                &       &        & 24.85   & 0.0102                & 0.00919     & 0.0111      \\
			\rowcolor{grey} &       & (W) 13 & 24.12   & 0.00978               & 0.00889     & 0.0107      \\
			\rowcolor{grey} &       &        & 25.44   & 0.00918               & 0.00833     & 0.0100      \\
			                &       & 14     & 29.64   & 0.00525               & 0.00471     & 0.00579     \\
			                &       &        & 31.33   & 0.00770               & 0.00706     & 0.00834     \\
			\rowcolor{grey} &       & (W) 15 & 39.11   & 0.00433               & 0.00390     & 0.00477     \\
			                &       & 16     & 41.05   & 0.00125               & 0.00100     & 0.00149     \\
			\rowcolor{grey} &       & 17     & 46.48   & 0.000780              & 0.000560    & 0.000999    \\
			                &       & (W) 18 & 49.88   & 0.000416              & 0.000196    & 0.000636    \\
			\rowcolor{grey} & Minor & 5      & 11.19   & 0.0340                & 0.0322      & 0.0358      \\
			\rowcolor{grey} &       &        & 12.17   & 0.0175                & 0.0153      & 0.0198      \\
			                &       & 6      & 13.35   & 0.0154                & 0.0132      & 0.0175      \\
			                &       &        & 14.25   & 0.0122                & 0.0110      & 0.0134      \\
			                &       &        & 15.23   & 0.00910               & 0.00825     & 0.00994     \\
			                &       &        & 16.16   & 0.00997               & 0.00895     & 0.0110      \\
			                &       &        & 17.00   & 0.00675               & 0.00543     & 0.00807     \\
			\rowcolor{grey} &       & 7      & 18.78   & 0.00445               & 0.00384     & 0.00506     \\
			\rowcolor{grey} &       &        & 19.95   & 0.00279               & 0.00229     & 0.00329     \\
			\rowcolor{grey} &       &        & 21.22   & 0.00311               & 0.00258     & 0.00363     \\
			                &       & 12     & 17.64   & 0.00684               & 0.00603     & 0.00765     \\
			                &       &        & 18.91   & 0.00543               & 0.00484     & 0.00601     \\
			                &       &        & 20.10   & 0.00435               & 0.00367     & 0.00503     \\
			\rowcolor{grey} &       & (W) 19 & 21.09   & 0.00268               & 0.00218     & 0.00317     \\
			\rowcolor{grey} &       &        & 22.55   & 0.00174               & 0.00133     & 0.00215     \\
			                &       & 20     & 26.92   & 0.00136               & 0.000932    & 0.00179     \\
			                &       &        & 27.97   & 0.00114               & 0.000826    & 0.00145     \\
			                &       &        & 29.25   & 0.00128               & 0.000881    & 0.00168     \\
			\rowcolor{grey} &       & 21     & 35.48   & 0.000588              & 0.000293    & 0.000883    \\
			                &       & 22     & 35.46   & 0.000786              & 0.000503    & 0.00107     \\
			                &       &        & 36.88   & 0.000536              & 0.000295    & 0.000778    \\
			\rowcolor{grey} &       & (W) 23 & 38.75   & 0.000333              & 0.000121    & 0.000546    \\
			                &       & 24     & 39.26   & 0.000336              & 0.000101    & 0.000571    \\
			                &       &        & 40.96   & 0.000195              & 0.000010    & 0.000410    \\
			\rowcolor{grey} &       & 26     & 44.34   & 0.000379              & 0.000181    & 0.000577    \\
			                &       & (W) 27 & 28.88   & 0.000789              & 0.000537    & 0.00104     \\
			\rowcolor{grey} &       & 28     & 29.45   & 0.00110               & 0.000864    & 0.00134
		\end{tabular} \label{Table_RawData} \\
	\end{center}
\end{table}
\begin{table}
\tiny
	\begin{center}
		\begin{tabular}{p{1.1cm} p{0.6cm} r r p{1.1cm} p{0.9cm} p{0.9cm}}
			\hline\hline
			Galaxy          & Axis  & Field  & r (kpc)     & N$\cdot$Arcsec$^{-2}$ & $-1 \sigma$ & $+1 \sigma$ \\ \hline
			NGC 4565        & Major & 7      & 27.62       & 0.112                 & 0.109084    & 0.115       \\
			                &       &        & 31.29       & 0.0420                & 0.0404      & 0.0436      \\
			                &       &        & 34.57       & 0.0260                & 0.0239      & 0.0280      \\
			\rowcolor{grey} &       & (W) 8  & 46.30       & 0.00397               & 0.00312     & 0.00483     \\
			\rowcolor{grey} &       &        & 48.77       & 0.00407               & 0.00326     & 0.00488     \\
			\rowcolor{grey} &       &        & 51.35       & 0.00474               & 0.00389     & 0.00560     \\
			\rowcolor{grey} &       &        & 53.57       & 0.00385               & 0.00296     & 0.00475     \\
			                & Minor & 1      & 8.98        & 0.114                 & 0.109       & 0.118       \\
			                &       &        & 10.36       & 0.0797                & 0.0758      & 0.0835      \\
			                &       &        & 11.72       & 0.0488                & 0.0456      & 0.0521      \\
			                &       &        & 13.04       & 0.0298                & 0.0246      & 0.0351      \\
			\rowcolor{grey} &       & 3      & 16.76       & 0.0200                & 0.0170      & 0.0231      \\
			\rowcolor{grey} &       &        & 18.05       & 0.0185                & 0.0164      & 0.0207      \\
			\rowcolor{grey} &       &        & 19.55       & 0.0119                & 0.0103      & 0.0134      \\
			\rowcolor{grey} &       &        & 21.03       & 0.00863               & 0.00734     & 0.00991     \\
			\rowcolor{grey} &       &        & 22.53       & 0.00683               & 0.00567     & 0.00799     \\
			\rowcolor{grey} &       &        & 24.18       & 0.00642               & 0.00530     & 0.00755     \\
			\rowcolor{grey} &       &        & 25.70       & 0.00545               & 0.00439     & 0.00650     \\
			\rowcolor{grey} &       &        & 27.15       & 0.00606               & 0.00478     & 0.00733     \\
			\rowcolor{grey} &       &        & 28.26       & 0.00460               & 0.00267     & 0.00654     \\
			                &       & (W) 11 & 34.95       & 0.00138               & 0.000801    & 0.001959    \\
			                &       &        & 37.96       & 0.00170               & 0.00122     & 0.00218     \\
			                &       &        & 41.31       & 0.000111              & 0.000010    & 0.000452    \\
			\rowcolor{grey} &       & 12     & 52.63       & 0.000486              & 0.000124    & 0.000849    \\
			\rowcolor{grey} &       &        & 56.57       & 0.00105               & 0.000696    & 0.00141     \\ \hline
			NGC 4945        & Major & (W) 2  & 20.93       & 0.0264                & 0.0246      & 0.0283      \\
			                &       &        & 21.99       & 0.0254                & 0.0238      & 0.0270      \\
			                &       &        & 23.06       & 0.0237                & 0.0220      & 0.0255      \\
			\rowcolor{grey} &       & 3      & 24.90       & 0.0229                & 0.0202      & 0.0255      \\
			\rowcolor{grey} &       &        & 25.95       & 0.0226                & 0.0210      & 0.0242      \\
			\rowcolor{grey} &       &        & 27.00       & 0.0213                & 0.0200      & 0.0226      \\
			\rowcolor{grey} &       &        & 28.12       & 0.0241                & 0.0225      & 0.0257      \\
			\rowcolor{grey} &       &        & 29.11       & 0.0200                & 0.0176      & 0.0225      \\
			                &       & (W) 4  & 25.72       & 0.0147                & 0.0134      & 0.0160      \\
			                &       &        & 26.81       & 0.0151                & 0.0138      & 0.0164      \\
			                &       &        & 27.82       & 0.0129                & 0.0116      & 0.0142      \\
			\rowcolor{grey} &       & 5      & 37.35       & 0.00687               & 0.00592     & 0.00782     \\
			\rowcolor{grey} &       &        & 38.75       & 0.00747               & 0.00673     & 0.00820     \\
			\rowcolor{grey} &       &        & 40.05       & 0.00533               & 0.00457     & 0.00609     \\
			                &       & (W) 6  & 44.41       & 0.00288               & 0.00217     & 0.00359     \\
			                &       &        & 45.46       & 0.00351               & 0.00282     & 0.00421     \\
			                &       &        & 46.63       & 0.00276               & 0.00203     & 0.00350     \\
			\rowcolor{grey} & Minor & (W) 7  & 8.92        & 0.0519                & 0.0482      & 0.0557      \\
			\rowcolor{grey} &       &        & 9.57        & 0.0413                & 0.0387      & 0.0439      \\
			\rowcolor{grey} &       &        & 10.29       & 0.0409                & 0.0383      & 0.0435      \\
			\rowcolor{grey} &       &        & 11.06       & 0.0328                & 0.0305      & 0.0352      \\
			\rowcolor{grey} &       &        & 11.70       & 0.0259                & 0.0234      & 0.0285      \\
			                &       & 8      & 12.14       & 0.0194                & 0.0178      & 0.0211      \\
			                &       &        & 12.92       & 0.0233                & 0.0216      & 0.0250      \\
			                &       &        & 13.79       & 0.0205                & 0.0189      & 0.0221      \\
			                &       &        & 14.63       & 0.0178                & 0.0164      & 0.0193      \\
			                &       &        & 15.40       & 0.0121                & 0.0108      & 0.0136      \\
			\rowcolor{grey} &       & (W) 9  & 15.23       & 0.00727               & 0.00552     & 0.00902     \\
			\rowcolor{grey} &       &        & 15.84       & 0.0127                & 0.0112      & 0.0142      \\
			\rowcolor{grey} &       &        & 16.63       & 0.0134                & 0.0120      & 0.0149      \\
			\rowcolor{grey} &       &        & 17.43       & 0.0108                & 0.00947     & 0.0121      \\
			\rowcolor{grey} &       &        & 18.05       & 0.00724               & 0.00561     & 0.00888     \\
			                &       & 10     & 18.63       & 0.00866               & 0.00748     & 0.00985     \\
			                &       &        & 19.39       & 0.00611               & 0.00517     & 0.00704     \\
			                &       &        & 20.30       & 0.00439               & 0.00356     & 0.00521     \\
			                &       &        & 21.14       & 0.00407               & 0.00327     & 0.00486     \\
			                &       &        & 21.89       & 0.00330               & 0.002500    & 0.00411     \\
			\rowcolor{grey} &       & (W) 11 & 33.28       & 0.000465              & 0.000094    & 0.000835    \\
			                &       & 12     & 40.24       & 0.000621              & 0.000264    & 0.000977    \\ \hline
			NGC 7814        & Major & 3      & 32.49       & 0.0249                & 0.0228      & 0.0269      \\
			                &       &        & 35.33       & 0.0185                & 0.0171      & 0.0199      \\
			                &       &        & 38.58       & 0.0175                & 0.0162      & 0.0189      \\
			                &       &        & 41.71       & 0.0131                & 0.0119      & 0.0143      \\
			                &       &        & 44.78       & 0.0133                & 0.0117      & 0.0148      \\
			\rowcolor{grey} &       & S4G    & 22.57$^{*}$ & 0.467$^{*}$           & 0.420$^{*}$ & 0.514$^{*}$ \\
			                & Minor & 4      & 19.31       & 0.0355                & 0.0335      & 0.0376      \\
			                &       &        & 22.83       & 0.0241                & 0.0226      & 0.0256      \\
			                &       &        & 26.79       & 0.0160                & 0.0148      & 0.0172      \\
			                &       &        & 30.33       & 0.00962               & 0.00857     & 0.0107      \\
			\rowcolor{grey} &       & S4G    & 8.95$^*$    & 0.860$^*$             & 0.770$^*$   & 0.930$^*$
		\end{tabular} \\
        $^*$Data point derived from the 3.6{\micron} S4G profile for 7814's axes fits.\\
        Points marked with a (W) are derived from WFC3 data; all other fields are from ACS.\\
        r (kpc) column shows radial distance from the galactic center.\\
	\end{center}
\end{table}

\begin{table}
\tiny
	\begin{center}
		\caption{Flux/Star Ratios for the Galaxies Examined in this Paper \label{tab:fluxstar}}  
        \begin{tabular}{cc}
        Galaxy&Flux Ratio (Absolute V-Band Mag$\cdot N^{-1}$) \\
        \hline
       NGC 253 &$2.06\cdot10^{-10}$\\
        NGC 891&$2.20\cdot10^{-10}$\\
        NGC 3031&$2.09\cdot10^{-10}$\\
        NGC 4565&$1.03\cdot10^{-10}$\\
        NGC 4945&$3.66\cdot10^{-10}$\\
        NGC 7814&$1.21\cdot10^{-10}$\\
		\end{tabular} \\
        Conversion from star density to apparent magnitude: $\mu_{v}=-2.5\cdot log_{10}(N\cdot Arcsecond^{-2} \cdot Flux Ratio)$
	\end{center}
\end{table}

\begin{table}
	\begin{center}
		\caption{Estimated `total' stellar halo mass fraction from Dragonfly imaging from the values presented in \citep{Merritt16}}
        \label{dragonfly} 
		\begin{tabular}{ll}
        \hline \hline
        Galaxy&Stellar halo fraction $f_{halo,tot}$ \\
        \hline
       NGC 1084& 0.15$\pm$0.07 \\
       NGC 2903& 0.031$\pm$0.021 \\
       NGC 3351& $<0.068$ \\
       NGC 3368& $<0.099$ \\
       NGC 4220& 0.027$\pm$0.022 \\
       NGC 4258 & $<0.057$ \\
       M101 & $<0.012$ \\
       \hline
		\end{tabular} \\
	\end{center}
\end{table}

\end{document}